\documentclass[12pt,twoside]{article}
\usepackage{correl}
\usepackage[version=4]{mhchem}  
\usepackage{cases} 
\usepackage{wasysym}
\usepackage{graphicx}
\usepackage[utf8]{inputenc}

\def\TheTitle{Mott Physics in Correlated Nanosystems: Localization-Delocalization Transition of Electrons by \textbf{E}xact \textbf{D}iagonalization \textit{\textbf{A}b \textbf{I}nitio} Method}
\def\TheHeading{Mott Physics in Correlated Nanosystems}
\def\TheAuthor{Józef Spałek}
\def\TheInstitute{Institute of Theoretical Physics, Jagiellonian University}
\def\TheAddress{ul. Łojasiewicza 11, PL-30-348 Kraków}
\def\TheChapter{5}           

\newcommand*{\eff}{\mathrm{eff}}
\newcommand*{\ep}{\epsilon}
\newcommand*{\de}{\delta}

\newcommand*{\abs}[1]{\left|#1\right|}
\newcommand*{\aver}[1]{\langle #1\rangle}
\begin{document}
\MakeTitle
\section{Introduction \& motivation: Localized versus itinerant.}

The textbook division of the electronic states in quantum  matter ranges between the two principal categories: \textit{(i)} localized (bound, atomic) states and \textit{(ii)} extended (delocalized, band, Fermi-liquid, free-particle-like) states. The two classes of states are depicted schematically in Fig. 1. While the question of existence of space-bound states in solids are described and characterized experimentally to a very good accuracy \cite{Griffith}, the transformation of those atomic states into emerging delocalized states in solid state physics (or in general, in condensed matter physics) is still under debate. The classical textbooks on the latter subject with a successful implementation of wave mechanics to molecular and metallic systems start with the Bloch theorem establishing the periodic nature of the single-electron states in solids under the influence of the corresponding translational symmetric single-particle potential. The first success  of the methods of the LCAO, H\"{u}ckel, etc. approaches was quite impressive given a total negligence of the interparticle interactions. Those interactions are not only of Coulomb type, but also of the e.g., van der Waals type which appear in molecular or solid-state systems. 

The question of including interpartical interactions in the context of periodic solid-state systems was posed qualitatively by Nevill Mott (for review see \cite{MottBook1990,Gebhard1997}). Mott based his argument on earlier experimental observations that for example the cobalt oxide \ce{CoO} that, according to the elementary Wilson classification of electronic states, should be regarded as a metallic system, since it possesses an odd number of valence electrons. Quite to the contrary, it was recognized as one of the best insulators known then. The argument was that probably the repulsive electron-electron interaction is responsible for a  destruction of coherent periodic Bloch states, as it favors  separating the particles from each other as far as possible, i.e., fixing them on the atoms states they originate from. Additionally, Mott argued later that the transition between atomic-type and itinerant (Bloch-type) of states should be discontinuous (first-order), since the Coulomb interaction is long-range, so the transition must take place from zero-concentration limit (insulting ground state) to the metallic state of sizable electron concentration, to warrant screening of the increasing - energy Coulomb interaction. In such a situation, those insulators should not only be clear-cut from metals with odd number of valence electrons, but also from the full-band Wilson-type insulators. In that, two features of such Mott (or Mott-Hubbard) systems should be singled out. First, as they contain unpaired spins, their magnetic ordering is tightly connected with them, usually of antiferromagnetic type as was discussed clearly by Anderson, see eg., \cite{AndersonPhysRevB1959} and Goodenough \cite{Goodenough}. Second, the Mott localization should be common to any condensed matter system, such as quark-gluon plasma \cite{Anderson_1963} or even cold-atom bosonic systems, and should appear if only the repulsive interparticle interaction is strong compared to their bare band (kinetic) energy. This shows a universal character of the related physics, particularly to those system in which  such a matter-insulator (localization-delocalization) transition is observed. A textbook of electronic system with such a transitions is the vanadium sesquioxide doped with chromium (\ce{(V_{1-x}Cr_x)_2O3} \cite{KuwamotoPhysRevB1980})

The aim of this chapter is to characterize first briefly the most striking properties of the Mott-Hubbard systems of macroscopic size and the  turn to the question of the localization in correlated nanosystems. The atomicity and itineracy of the valence electrons in the latter case can then be seen clearly on example of exact results, at least for the model systems. This analysis should provide us with additional arguments for the Mott phenomenon universality within the physics of quantum condensed matter. 

    \begin{figure} 
    \centering
   \includegraphics[width=0.7\textwidth]{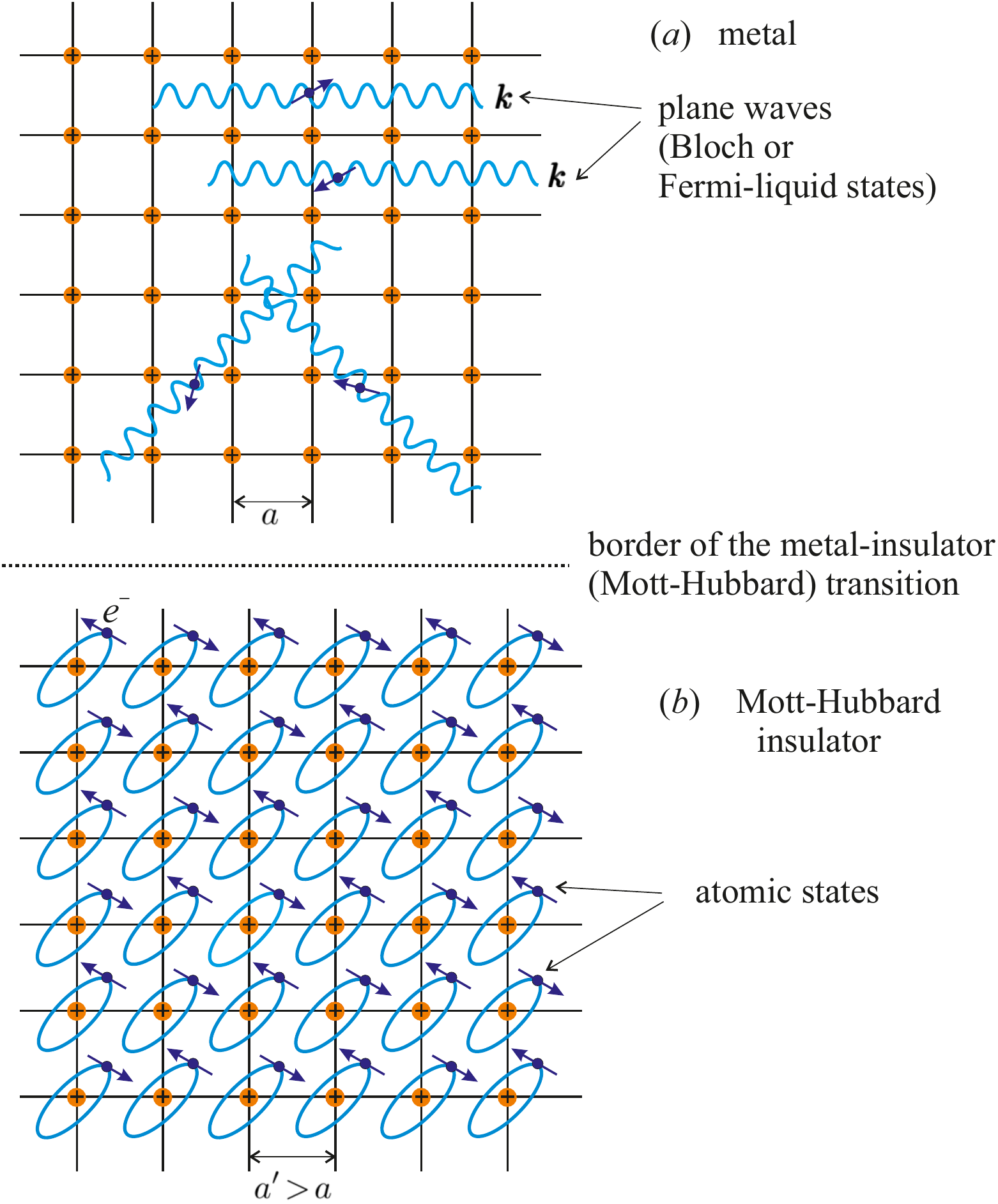}.
    \caption{Schematic representation of metallic (Fermi-liquid) (a) and Mott-Hubbard \index{Mott-Hubbard} insulating  (localized) states (b). Note that in the state (a) electrons derive from the parent atoms, which form a background lattice of cations (red solid points). The spins of unpaired electrons form as a rule an antiferromagnetic lattice.}
    \label{Fig_1}
  \end{figure}

  \begin{figure}  
    \centering
   \includegraphics[width=\textwidth]{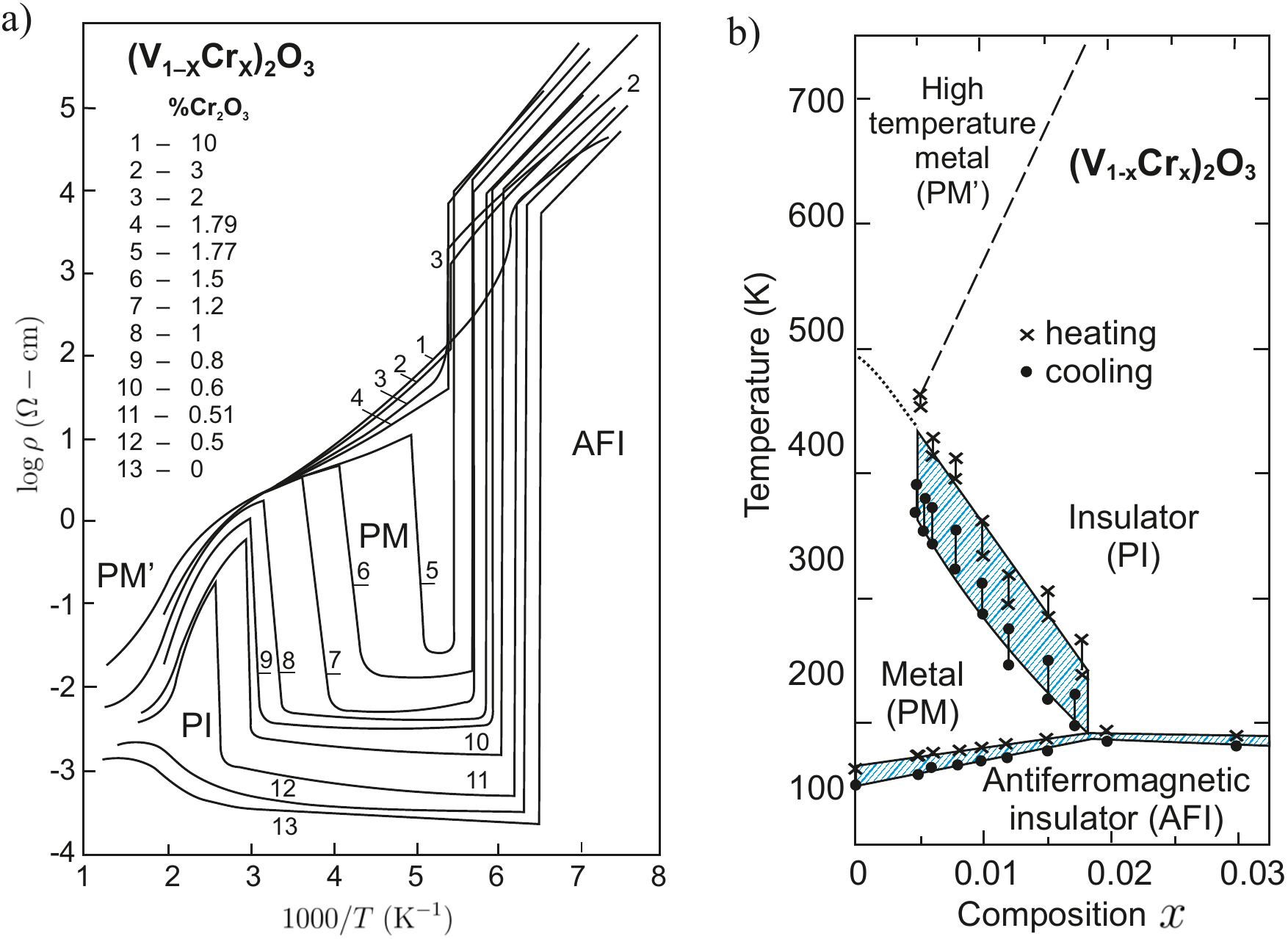}
\caption{ a) temperature dependence of the electrical resistivity (in logarithmic scale) vs. $1/T$ for Cr-doped $\mathrm{V_2}\mathrm{O_3}$. A very sharp transition from antiferromagnetic insulating (AFI) to paramagnetic metallic (PM) phase is followed by a reverse PM $\rightarrow$ PI at higher temperature, which in turn is followed by PI $\rightarrow$ PM' crossover transition to a reentrant metallic (PM') phase at still higher temperatures; b) phase diagram for the same system on $T$-$x$ plane; the hatched are depicts the hysteretic behavior accompanying the discontinuous  transitions (taken from Ref. \cite{KuwamotoPhysRevB1980, SpaekPhysRevLett1990}, with small modifications. Both AFI $\rightarrow$ PM and PM $\rightarrow$ PI represent examples of the Mott-Hubbard transition (see main text).}
\label{Fig_2}
  \end{figure}

\section{Essence of Mott-Hubbard localization: A physical picture}

In this section we define the concept of almost localized Fermi liquid
and the thermodynamic character of the Mott-Hubbard transition for electrons in a single narrow band. This picture is based on the Hubbard model and its direct variational analysis. 

\subsection{Definitions}


The ground-state energy of a periodic system of fermions can be described by starting from the system atomic configuration and, subsequently, adding other dynamic interactions which appear in the emerging condensed state. Namely, its energy per atomic state can be simply expressed in the form of \cite{SpalekEurJPhys2000}

  \begin{align}
    \frac{E_G}{N} = \epsilon_a + \langle T \rangle + \langle V \rangle + \langle V_{12} \rangle \equiv E_1 + E_2,
  \end{align}

  \noindent
  where $\epsilon_a$ is the single particle energy in an atomic (\emph{Wannier}) state, $\langle T\rangle$ and $\langle V\rangle$ are the average kinetic and potential energies in, whereas $\langle V_{12}\rangle$ is the expectation value of the two-particle interaction. The single-particle part $E_1$ comprises the first three terms, and $E_2 \equiv \langle V_2\rangle$. In such a periodic system near the delocalization–localization transition, we usually assume
that $\epsilon_a = 0$; i.e., it is regarded as a constant (reference) value which is often disregarded unless stated explicitly (see next Sections). In this manner, the remaining terms characterize solely the energy contributions of relevant fermions in condensed state with respect to that in the atomic state. Note also that usually $E_1 < 0$. Next, one can define two physically distinct regimes:

  \begin{enumerate}
  \item [\emph{1}$^{\circ}$] $|E_1| \gtrsim E_2$: Fermi-liquid (metallic) regime \index{Fermi-liquid regime}, ranging from a simple-metal region ($|E_1| \gg E_2$), through the  Fermi-liquid regime, to the delocalization-localization threshold when $(E_{1})\approx E_2$;
  \item [\emph{2}$^{\circ}$] $|E_1| \ll E_2$: Strong-correlation (Mott-Hubbard) regime. \index{Mott-Hubbard} 
  \end{enumerate}

Let us characterize briefly each of them and introduce the states in these regimes. Connected with this we  start from atomic (Wannier) representation of the involved states and interactions, in the situation \emph{1}$^{\circ}$. The starting point is described then by either a gas of fermions or the Landau Fermi liquid, and associated with both of them momentum representation and the Fermi-Dirac statistics (distribution) in its canonical form. In discussing the correlated system, we start as a rule from the Wannier representation (see below). This means that, in general, we can start from two \emph{complementary} representations of the  single-particle quantum-mechanical states, i.e., either from the Bloch representation, in which the momentum uncertainty is zero, or from the Wannier representation, in which the proper quantum number characterizing the state is a fixed lattice position, at which the wave function is centered.   
The above division into the two asymptotic regimes $|E_1| \gg E_2$ and $|E_1| \ll E_2$ is illustrated in Fig.\ref{Fig_1}, where the \emph{complementary} nature of the single-particle states is represented on example of a solid with metallic (delocalized) states of electrons (a) or correlated (atomic, Mott) states (b) for the case with one relevant valence electron per parent atom. Additionally, we have marked a dividing line (\emph{the Mott-Hubbard boundary}) between the two macrostates. The momentum representation is described by set of the Bloch functions $\left\{\Psi_{\mathbf{p}\sigma}(\mathbf{r})\right\}$ with (quasi)momentum $\mathbf{p} = \hbar \mathbf{k}$ and the spin quantum number $\sigma = \pm 1 \equiv\, \uparrow, \downarrow$, whereas the position representation is expressed by the corresponding set of Wannier states $\left\{ w_{i\sigma}(\mathbf{r})\right\}$. These two representations are equivalent in the sense that they are related by the lattice Fourier transformation. However, in the situation depicted in Fig.\ref{Fig_1}, when we have a sharp boundary (usually \emph{ first-order phase-transition line}) between the states shown in (a) and (b), this equivalence is broken and, in effect, the unitary symmetry U($N$) does not apply. The macroscopic state (a) near the transition is represented, strictly speaking, by a modified Landau-Fermi liquid (the so-called \emph{almost localized Fermi liquid}, ALFL), whereas the Mott-insulating state is well accounted for as a localized-spin (Heisenberg) antiferromagnet. \index{almost localized Fermi liquid}

From the above qualitative picture one can infer that with approaching \emph{metal $\rightarrow$ insulator boundary}, i.e., with formation of the localized-spin state, the kinetic energy of the \emph{ renormalized-by-interaction} particle progressive motion throughout the system is drastically reduced and, as a result, it reduces to zero in the localized (insulating) state. Effectively, one can say that then the Landau quasiparticle effective mass $m^{*} \rightarrow \infty$. This feature illustrates the situation that strong enough interactions (called in this context \emph{strong correlations}) limit the stability of the Landau-Fermi quasiparticle picture, as is exemplified explicitly by the appearance of the Mott-Hubbard phase transition. Also, a proper quantitative description of the transition requires a model with a simultaneous generation of the effective exchange interactions (\emph{kinetic exchange} \cite{KochBook1988} in the one-band case or superexchange in the multi-orbital situation). In the subsequent Sections 2 and 3 we provide a quantitative analysis of these statements. The starting point of these considerations is the parametrized microscopic Hamiltonian provided below. We limit our discussion to the Hubbard model as an illustration of more complicated analysis of nanophysical systems in Sections 4 and 5. \index{Mott-Hubbard phase transition} \index{Hubbard model} \index{Mott insulators}

\subsection{Correlated (nano)materials}

The following examples of bulk systems belonging to \emph{1}$^{\circ}$ or \emph{2}$^{\circ}$

  \begin{enumerate}
  \item [\emph{1}$^{\circ}$] Mott-Hubbard systems: (\ce{(V_{1-x}Cr_x)_2O3} \cite{KuwamotoPhysRevB1980}), \ce{NiS_{2-x}Se_x} 
  \cite{HonigSpalek}, organic metals \cite{KanodaKato}.
  \item [\emph{2}$^{\circ}$] Mott (antiferromagnetic) insulators: \ce{NiO}, \ce{CoO}, \ce{La_2CuO_4}, \ce{YBa_2Cu_3O_7}, etc. Strongly correlated metals: high temperature superconductors: \ce{La_{2-x}Sr_xCuO_{4}}, \ce{YBa_2Cu_3O_{7-\delta}}, heavy fermion systems: \ce{CeAl_3}, \ce{CeCu_2Si_2}, \ce{UBe_{13}}, \ce{CeCoIn_5}, etc. 
  \end{enumerate}
  
These are the most typical systems and can be regarded as almost localized Fermi-liquids. 
There are also systems with quantum phase transitions and non-Landau (non-Fermi) liquid states, but those are regarded as a separate class as then the quantum fluctuations are as important on the correlations. Those systems are not tackled in detail here. 

At this point we would like to say few words about the correlated nanosystems. The atomic (bound) states are localized by definition. The basic question is what happens when we form e.g., a nanochain or nanoring. How such a small system  can become a nanometal (e.g., a monoatomic quantum wire), when we vary the interatomic distance? In other wards, at what point the set of discrete atomic states form a nanoliquid? We address this type of questions after analyzing first the nature of the delocalization in bulk systems. 

\subsection{From Landau-Fermi liquid to Mott-Hubbard insulator through an almost localized Fermi liquid}
\index{Landau-Fermi liquid} \index{quasiparticle}

The Landau theory of Fermi liquids represents a standard reference point in the theory of interacting fermions (for recent references see \cite{BaymBook1991,Hindus,SpalekPhysRevB1990}).
Here we characterize only briefly their characteristics, particularly those which appear or are relevant to theory of correlated systems.

The principal assumption of the theory is that we are interested  in the changes of \emph{ideal-Fermi-gas-properties}, which are induced by the inter-particle interactions and associated with them thermal excitations at low temperatures. In other words, we express the change of the total energy of the system due to the appearing interaction in the Landau form
\begin{equation}
    \delta E \simeq \sum_{\pmb{k}\sigma} \epsilon_{\pmb{k}\sigma}\, \delta n_{\pmb{k}\sigma} + \frac{1}{2} \sum_{\pmb{k}\pmb{k}'} f^{\sigma\sigma'}_{\pmb{k}\pmb{k}'}\, \delta n_{\pmb{k}\sigma}\, \delta n_{\pmb{k}'\sigma'} \equiv \sum_{\pmb{k}\sigma} E_{\pmb{k}\sigma} \delta n_{\pmb{k}\sigma},
\end{equation}
where $\epsilon_{\pmb{k}\sigma}$ is the single-particle energy (with respect to the chemical potential terms $\mu$) and $f^{\sigma\sigma'}_{\pmb{k}\pmb{k}'}$ (generally spin-dependent) is the effective interaction between those particles; it has the form of spin-dependent density-density interactions. Explicitly, the bare-particle energy in the Zeeman field $H_{a}$ is $\epsilon_{\pmb{k}\sigma} \equiv \epsilon_{\pmb{k}} -g\mu_{B}H_{a}\sigma-\mu$ and in the isotopic liquid (not generally true for fermions in lattice systems) we have that $f^{\sigma\sigma'}_{\pmb{k}\pmb{k}'} = f^{s}_{\pmb{k}\pmb{k}'} (\pmb{k}\cdot\pmb{k}'/k^{2}_{F}) + \sigma\sigma'f^{a} (\pmb{k}\cdot\pmb{k}'/k^{2}_{F})$, where $k_{F}$ is the Fermi wave vector and $f^{s,a}$ express spin-independent and spin-dependent parts, respectively. The next assumption is that we take into account the interaction-induced scattering processes for particles at the Fermi surface, i.e, put that
$\frac{\pmb{k}\cdot\pmb{k}'}{k^{2}_{F}}= \cos{\theta_{\pmb{k}\pmb{k}'}}$
and subsequently we can express the interaction parameters in terms of Legendre polynomial expansion
\begin{equation}
    f^{(s,a)}(\cos{\theta})= \sum^{\infty}_{l=0} f^{(s,a)}_{l} P_{l}(\cos{\theta}).
    \label{r_2.2}
\end{equation}
There are three basic assumptions in the Landau formulation of the Fermi-liquid theory. First, the interparticle scattering is important only very near or, strictly speaking, at the Fermi surface due to the Pauli principle, i.e., the circumstance that particles can scatter only from occupied states $|\pmb{k}\sigma\rangle$ into unoccupied ones. Second, a well defined Fermi surface remains intact even if the scattering processes are included ( this is \emph{the Luttinger theorem} proved later on the grounds of perturbation expansion and assuming validity the Dyson theorem validity then this is not always valid for correlated systems). Third, there is \emph{one-to-one} correspondence between the initial (bare energy states, $\epsilon_{\pmb{k}\sigma}$) and the effective (\emph{quasiparticle}) states with energies $E_{\pmb{k}\sigma}\equiv \epsilon_{\pmb{k}\sigma} + \frac{1}{2}\sum_{\pmb{k}\pmb{k}'\sigma'} f^{\sigma\sigma'}_{\pmb{k}\pmb{k}'} \delta n_{\pmb{k}'\sigma'}$.
Moreover, the Fermi energy value $E_{F}\equiv \mu$ at $T=0$ can be regarded as the reference energy for both the \emph{bare-} and \emph{quasi-particle} states. Effectively, this means that the interaction processes, practically active only at the Fermi surface, do not influence the Fermi surface volume. Finally, from the third assumption it follows that the statistical distribution for the quasiparticles can be taken in the form of the Fermi-Dirac distribution for those states, i.e.,  $f(E_{k\sigma})=[\exp{(\beta E_{k\sigma})}+1]^{-1}$.

\begin{figure}    
    \centering
    \includegraphics[width=\textwidth]{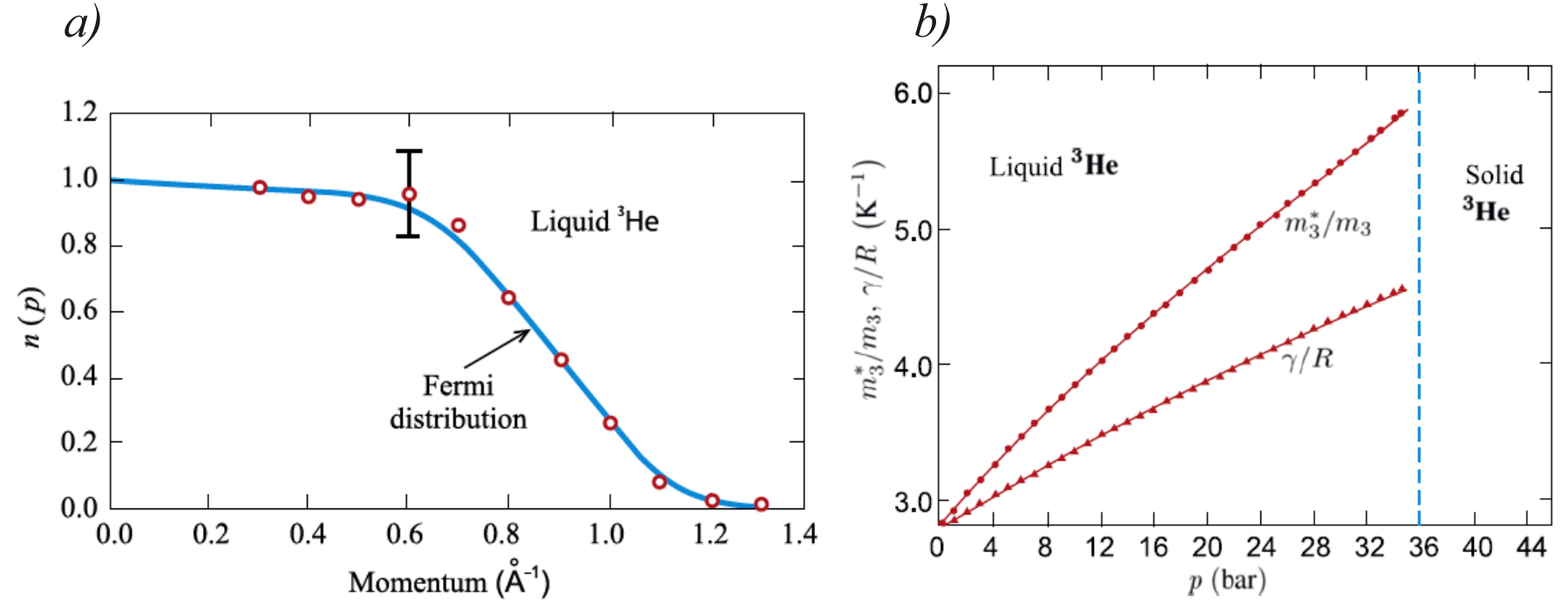}
    \caption{Principal characteristics of liquid \ce{^3He} as of Fermi liquid: a) the Fermi-Dirac distribution measured by neutron scattering at ambient pressure and temperature $T=0.37$K \cite{MookPhysRevLett1985}; \index{Fermi liquid}
    b) the linear-specific-heat coefficient $\gamma$ in units of gas constant $R$ and inferred from it effective atom mass enhancement $m^{*}_3/m_3$, both as a function of external pressure \cite{GreywallPhysRevB1986}.
    The vertical dashed line marks the liquid-solid transition, regarded in this case as discontinues Mott transition to the localized state of whole atoms. The spin $1/2$ is attached to the nuclei for this case two-electrons atoms in $1s$$^2$ configuration.}
    \label{fig_2_1}
  \end{figure}

The additional ingenious feature of this theory is the circumstances that the principal properties of the Fermi liquid, such as liquid \ce{^3He}, can be expressed solely by the first three parameters of expansion (\ref{r_2.2}): $f^{s}_{0}$, $f^{s}_{1}$, and $f^{a}_{0}$, what makes this theory, even though phenomenological in its nature, fully testable in its original form, at least for the isotropic quantum liquid \ce{^3He}. What is more important, the assumption about the Fermi-Dirac distribution applicability has been tested on two systems: experimentally, for liquid \ce{^3He} (cf. Fig. \ref{fig_2_1}ab) and theoretically by considering evolution of the statistical distribution function, calculated exactly for a model nano-chains and nano-rings of hydrogen atoms, as a function of interatomic distance \cite{RycerzPhDThesis,SpalekJPCM2007} (see later here).

Nevertheless, as shown in Fig.~\ref{fig_2_1}a and b, the effective-mass concept ($m^{*}_{3}$) for 
\ce{^3He} atoms breaks down and consequently, of the linear specific heat $\gamma$ ceases the exist, at the liquid-solid transition (cf. Fig. \ref{fig_2_1}b). These effects cannot be accounted within 
the Landau-Fermi liquid theory. We discussed that question next within the Hubbard model by introducing first the concept of \emph{an almost localized Fermi liquid} and as a consequence, a discontinuous delocalization-localization (metal-insulator) phase transition. These aspects are regarded as the fundamental features of correlations.

  \index{Landau-Fermi liquid}
  \index{almost localized Fermi liquid}

\subsection{The concept of almost localized Fermi liquid (ALFL)}

One can notice from Fig. \ref{fig_2_1}b that the Fermi-liquid state characterized there by the linear specific-heat coefficient (in unite of gas constant $R$), $\gamma/R$ and the resulting from it effective-mass enhancement $m^{*}_{3}/m_3$ of the \ce{^3He} atom in this milieu, both loose their meaning at \emph{the liquid-solid transition}, which takes place at the relatively low external pressure $\simeq 36$ bar. At this point the atoms freeze into well-defined crystal positions and their individual quantum mechanical states are characterized from now on by set of Wannier functions $\{w(\textbf{r}\cdot\textbf{r}_{i})\}$ centered at well-defined lattice sites $\{\textbf{r}_{i}\}$.
It must be underlined that in this case there is no external single-particle potential trapping the particles, as it is the case of electrons in solids. Such a solidification is regarded thus as an example of a spontaneous breakdown of transitional symmetry, albeit in a discontinuous manner.  
Our task in this Section is to briefly discuss the delocalization states in the metallic liquid of electrons close to the transition to the localized state, and next, explain its first-order phase-transition nature. \index{Hubbard model}

We model the system by starting from the Hubbard Hamiltonian (\ref{r_3}) and calculate first the system ground-state energy per atomic site $\langle \mathcal{H}\rangle/N$. 
The interaction between correlated particles in the simplest form is taken in the form of single-band Hubbard model \cite{HubbardProcRoySocA1963,HubbardProcRoySocA1964} with $\epsilon_{a}=0$ (i.e., $t_{ii}=0$; hence the primed summation in the first term),
\begin{equation}
    \mathcal{\tilde{H}} = {\sum_{ij\sigma}}' t_{ij}\,\hat{a}^{\dag}_{i\sigma} \,\hat{a}_{j\sigma} + U \sum_{i} \hat{n}_{i\uparrow}\,\hat{n}_{i\downarrow},
    \label{r_3}
  \end{equation}
in which $t_{ij}\equiv \langle w_{i}|\mathcal{H}_{1}|w_{j}\rangle <0$ represents the single-particle parameter phrased as \emph{the hopping parameter}, (and with the bandwidth of bare states $W\equiv 2z\left|\sum_{j(i)}t_{ij}\right|$,where $j(i)$ means the summation over neighboring sites to $i$), and $U$ is the magnitude of intraatomic interactions the so-called \emph{Hubbard term}. For strongly correlated electrons we can rephrase the conditions \emph{1}$^{\circ }$ and \emph{2}$^{\circ }$. Namely, situations with $W\apprle U$ or $W\simeq U$ represent the systems below and at the Mott-Hubbard transition, respectively, whereas the $W\ll U$ case represents the strongly-correlation limit \emph{2}$^\circ$. 
Note again that the primed summation in (\ref{r_3}) excludes the $i=j$ term $\sum_{i\sigma}t_{ij}\hat{n}_{i\sigma}=t_{0}N_e$, when the system is transitionally invariant ($t_{ii}=t_{0}\; \forall i$); then, $N_e=\sum_{i\sigma}\hat{n}_{i\sigma}$ is the total number of particles of $N$ atomic sites ($n\equiv N_e/N$ is the so-called band filling). If we regarded that the reference atomic energy of each of the electrons does not change near the metal-insulator transition ($W\simeq U$), then to $N_0$ can be thought of an irrelevant constant term (reference energy) and disregarded. This assumption must be revised (see later) as one includes an ab initio calculations, i.e., when the parameters are also calculated explicitly. But, first we analyze the situation as a function of $U/W$ for the half-filled ($n=1$) situation.

When approaching the localization-delocalization transition we expect that the single-particle and interaction parts become of comparable amplitude. Due to this circumstance, we assume that the hopping probability $\langle \hat{a}^{\dag}_{i\sigma}\, \hat{a}_{j\sigma} \rangle$ is renormalized by the interaction to the form $\langle \hat{a}^{\dag}_{i\sigma}\, \hat{a}_{j\sigma} \rangle \equiv q\langle \hat{a}^{\dag}_{i\sigma}\, \hat{a}_{j\sigma} \rangle_{0}$, where
$\langle \hat{a}^{\dag}_{i\sigma}\, \hat{a}_{j\sigma} \rangle_{0}$ is the hopping probability for noninteracting  \emph{(uncorrelated)} particles and $q$ is the so-called renormalization (band narrowing) factor: $q\rightarrow 1$ when $U\rightarrow 0$ and $q\rightarrow 0$ when $U\rightarrow U_C$, where $U_C$ is the critical interaction value for the transition to the localized state to the take place. Explicitly, we can write down the system internal energy in the form (for $U\leqslant U_C$) \cite{SpalekPhysRevLett1987,SpalekPhysRevB1989}
\begin{equation}
    \frac{E_G}{N}= \frac{1}{N}\sum_{\pmb{k}\sigma} E_{\pmb{k}}\, f(E_{\pmb{k}}) +U d^2,
\end{equation}
where $E_{\pmb{k}}\equiv q \epsilon_{\pmb{k}}$, $d^2 \equiv \langle \hat{n}_{i\uparrow}\, \hat{n}_{i\downarrow} \rangle$, and $f(E_{\pmb{k}})$ in the Fermi-Dirac function for renormalized particles regarded still as quasiparticles. In this expression $d^2$ is regarded as a variational parameter to be calculated self-consistently. Therefore, the whole problem reduces to determining microscopically $q\equiv q(d^2)$. This can be carried out by considering Gutzwiller variational approach \cite{BrinkmanPhysRevB1970}, but also from physical considerations \cite{SpalekPhysRevB1983}. It turns out that for the half-filled ($n=1$) state (i.e., with one particle per atomic site) and for systems with electron-hole symmetry this factor can be calculated in the elementary manner \cite{SpalekPhysRevB1983}
which yields simple result $q(d^2)= 8d^2(1-2d^2)$. Additionally, we have that for a constant density of states, the chemical potential we can be set $\mu \equiv 0$ and thus for $H_{a}=0$ we have \index{Gutzwiller variational approach}

\begin{equation}
    \bar{\epsilon} \equiv \frac{1}{N} \sum_{\pmb{k}\sigma} \left( E_{\pmb{k}}/q \right) = -\frac{W}{4},
    \label{r_2.3}
\end{equation}
where $E_{\pmb{k}}/q \equiv \epsilon_{\pmb{k}}$ represents, as before, the single particle energy of bare particles at the temperature $T=0$; also, the effective-mass renormalization is $m^{*}=m_{B}/q$, where $m_{B}$ is the bare band mass.

By minimizing energy (\ref{r_2.3}) with respect to $d^2$ we obtain both the physical ground-state energy and the quasiparticle energy spectrum $\{ E_{\pmb{k}} \}$. This in turn, allows us to calculate concrete ground-state and thermodynamic
properties. Explicitly  \cite{SpalekPhysRevB1989,SpalekEncyclopedia2005,SpalekRefModuleMateSci2016},

\begin{numcases}{}
d^{2} = \frac{1}{4} \left( 1-\frac{U}{U_C} \right),
\label{r_2.4}
\\
\frac{E_G}{N} = \frac{1}{4} \left( 1-\frac{U}{U_C} \right)^{2} \bar{\epsilon}\, ,
\label{r_2.5}
\\
\frac{m^{*}}{m_{0}} = \frac{1}{1-\left(\frac{U}{U_C} \right)^{2}} \equiv \frac{1}{q_{0}} \equiv 1+\frac{1}{3} F_{1}^{s}\, ,
\label{r_2.6}
\\
\gamma=\gamma_0 \frac{m^{*}}{m_{0}}= \gamma_0 \frac{1}{q_0}= \gamma_0 \left( 1+\frac{1}{3} F_{1}^{s}\right)\,,
\\
\chi=\chi_0\,\frac{1}{q_0 \left[ 1-\rho_0(\mu)\,U\cdot \frac{1+U/2U_c}{\left(1+U/U_c \right)^2} \right] } \equiv
\chi_0\,\frac{m_{0}}{\left(m^{*}\right) \left(1+F_{a}^{0}  \right)}\,,
\\
\frac{\chi}{\gamma} = \frac{\chi_0}{\gamma_0}\,  \frac{1}{  1-\rho_0(\mu)\,U\cdot \frac{1+U/2U_c}{\left(1+U/U_c \right)^2}  }\, ,
\label{r_2.7}
\end{numcases}

with $U_C\equiv 8|\bar{\epsilon}|=2W$ (the second value is for constant density of states). Additionally, to calculate the magnetic susceptibility $\chi$, a full Gutzwiller approach have been used \cite{BrinkmanPhysRevB1970}. When $U \rightarrow U_C \rightarrow 0$, $d^2 \rightarrow 0$, the ground-state energy $E_G \rightarrow 0$, the effective mass $m^{*} \rightarrow \infty$, and the magnetic susceptibility to linear specific heat coefficient $\chi /\gamma \rightarrow 4$. We see that at the transition, the interaction ($>0$) and the single particle ($<0$) parts compensate each other, the mass for a translational motion throughout the system diverges, and the magnetic susceptibility is roughly proportional to $\gamma$. The $U=U_C$ point thus represents a dividing line between the itinerant and atomic states of the matter and the freezing of particles into a lattice breaks the whole system translational invariance (at least, in liquid \ce{^3He} case). A full microscopic approach requires an explicit determination of the parameters $U$ and $\bar{\epsilon}$ as a function of pressure. Low-temperature corrections to above (\ref{r_2.4})-(\ref{r_2.7}) have been detailed elsewhere \cite{SpalekPhysRevB1989,SpalekPhysRevB1990}. The expression appearing on r.h.s. of the second quality signs represent the results from the Landau theory.  \index{Gutzwiller approach}

One may say that the picture formed by the expressions (\ref{r_2.4})-(\ref{r_2.7}) represents, as in any Fermi-liquid theory, a basic quasiparticle picture, with the additional boundary of its applicability for $U<U_C$. In fact, this picture can be mapped into the Landau-Fermi-liquid parametrization of the physical properties at $T=0$  \cite{VollhardtRevModPhys1984}. The question remains what are the collective spin- and charge-excitation spectra in the present case. This subject is a matter of our present studies and will not be detailed here \cite{FidrysiakPreprint2019,FidrysiakPhysRevB2017}.

\begin{figure}   
    \centering
    \includegraphics[width=\textwidth]{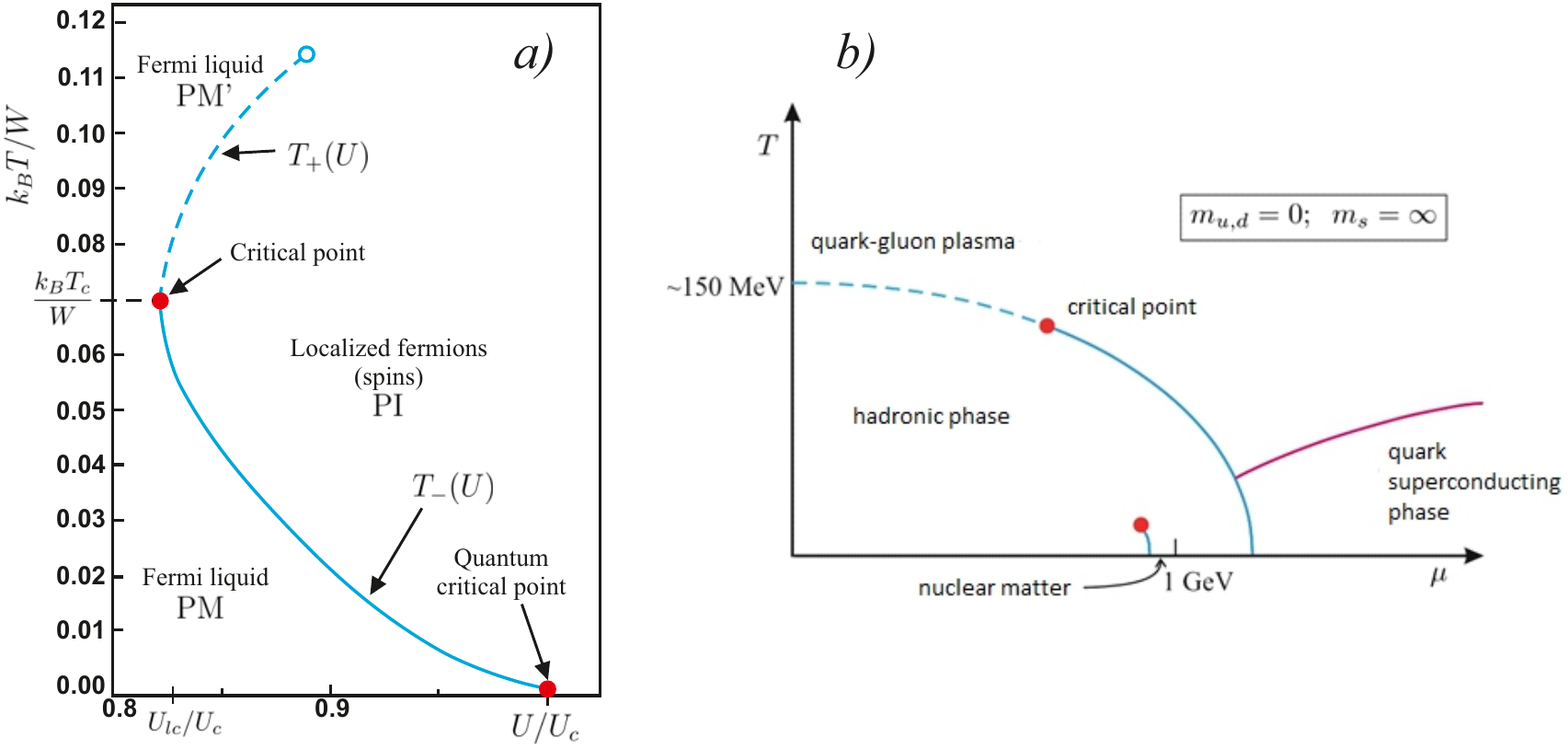}
    \caption{a) Phase diagram at $T\neq 0$ for almost localized fermions on the plane temperature $T$ versus relative interaction magnitude $U/U_C$. Note the presence of two critical points: classical at $T=T_c$ and quantum at $T=0$. This phase diagram does not include the magnetic phases (see below \cite{SpalekPhysRevLett1987}); b) an analogical phase diagram for the nuclear matter \cite{RajagopalNuclPhysA1999}. 
    The dashed lines represent extrapolations to high-temperature regime in both cases.}
    \label{fig_2_2}
  \end{figure}

\subsection{Delocalization-localization (Mott-Hubbard) transition}

As has been mentioned in the preceding Section, the  delocalization-localization transition at $T=0$ takes place at $U=U_C \approx W$. The question is when this transition will appear at arbitrary $T \geqslant 0$. This question is a nontrivial one, since near the transition, the renormalized single-particle and interaction energy not only almost compensate each other, but also each of the two terms vanishes separately. In such a situation, small perturbations such as the thermal or atomic disorder, applied magnetic field, or even the onset of magnetic order may balance out two quantum-mechanical contributions towards either insulating (localized) or itinerant (ALFL, metallic) state. We discuss the effect of nonzero temperature. \index{Landau-Fermi liquid}

Starting from the internal energy (\ref{r_2.2}) we define now the free energy functional of the itinerant correlated system  \cite{SpalekPhysRevLett1987,SpalekPhysRevB1989} as follows
\begin{equation}
    \frac{\mathcal{F}}{N}=\frac{1}{N}\sum_{\pmb{k}\sigma} E_{\pmb{k}}\,f_{\pmb{k}\sigma} +Ud^2+\frac{k_BT}{N}\sum_{\pmb{k}\sigma} \left[ f_{\pmb{k}\sigma}\ln{f_{\pmb{k}\sigma}} +(1-f_{\pmb{k}\sigma}) \ln{(1-f_{\pmb{k}\sigma})} \right],
    \label{r_2.8}
\end{equation}
where $f_{\pmb{k}\sigma}$ the Fermi-Dirac function for quasiparticles with energies $E_{\pmb{k}\sigma}$ and the last term is the entropy in the given, not necessarily, the  equilibrium state, which we determine subsequently by minimizing $\mathcal{F}$. This expression allows also for developing the low-temperature (Sommerfeld-type) expansion defined as the regime with $k_BT/q W\ll 1$.  In effect, the first non-trivial terms in paramagnetic state have the form
\begin{equation}
\frac{\mathcal{F}}{N} = -q\frac{W}{4} + Ud^2-\frac{\gamma_0T^2}{q} +O(T^4).
\end{equation}
After a minimizing of the functional $\mathcal{F}$ with respect to the $d^2$ we obtain the physical free energy $\mathcal{F}$ of ALFL. A detailed analysis of the low-$T$ expansion is provided  in \cite{SpalekPhysRevB1989}, where the Gutzwiller-Brinkman-Rice approach is generalized to $T>0$ case. Note that the expressions describe the free energy functional for an almost localized  Fermi liquid to be minimized with respect to $d^2$. As before, we assume that $\mu\equiv 0$, which means that the electron-hole symmetry holds.
The next step is to introduce the concept of discontinuous phase transition in the context of this fermion itinerant state instability. We regard the ALFL as a well-defined phase in the thermodynamic sense and the lattice of localized electrons (spins) as the other. Then, the discontinuous phase boundary between them is determined from the coexistence condition $F=F_I$, where $F_I$ is the free energy of the insulating state and has a very simple form if the spins are disordered
\begin{equation}
    \frac{F_I}{N}=-k_BT\,\ln{2},
\end{equation}
where $k_B\,\ln{2}$ is the entropy of $S=1/2$ spins.  From the coexistence condition, we obtain two transition temperatures
\begin{equation}
    k_BT_{\pm}=\frac{3q_0}{2\pi^2}\,W \left\{\ln{2}\pm \left[ (\ln{2})^2-\frac{\pi^2}{3}\left(  1-\frac{U}{U_C}\right)^2\!\Bigg{/}q_0 \right]^{1/2}
    \right\}.
    \label{r_2.11}
\end{equation}
The two solutions coalesce at $T_{+} \equiv T_{-}=T_c$ for $U=U_{lc}$, i.e., for the lowest critical value of the interaction for the transition to take place, which is determined from the condition

\begin{equation}
    \frac{U_{lc}}{U_C}=1-\frac{\sqrt{3}\,\ln{2}}{\pi}.
    \label{r_2.12}
\end{equation}

\begin{figure}   
    \centering
   \includegraphics[width=0.7\textwidth]{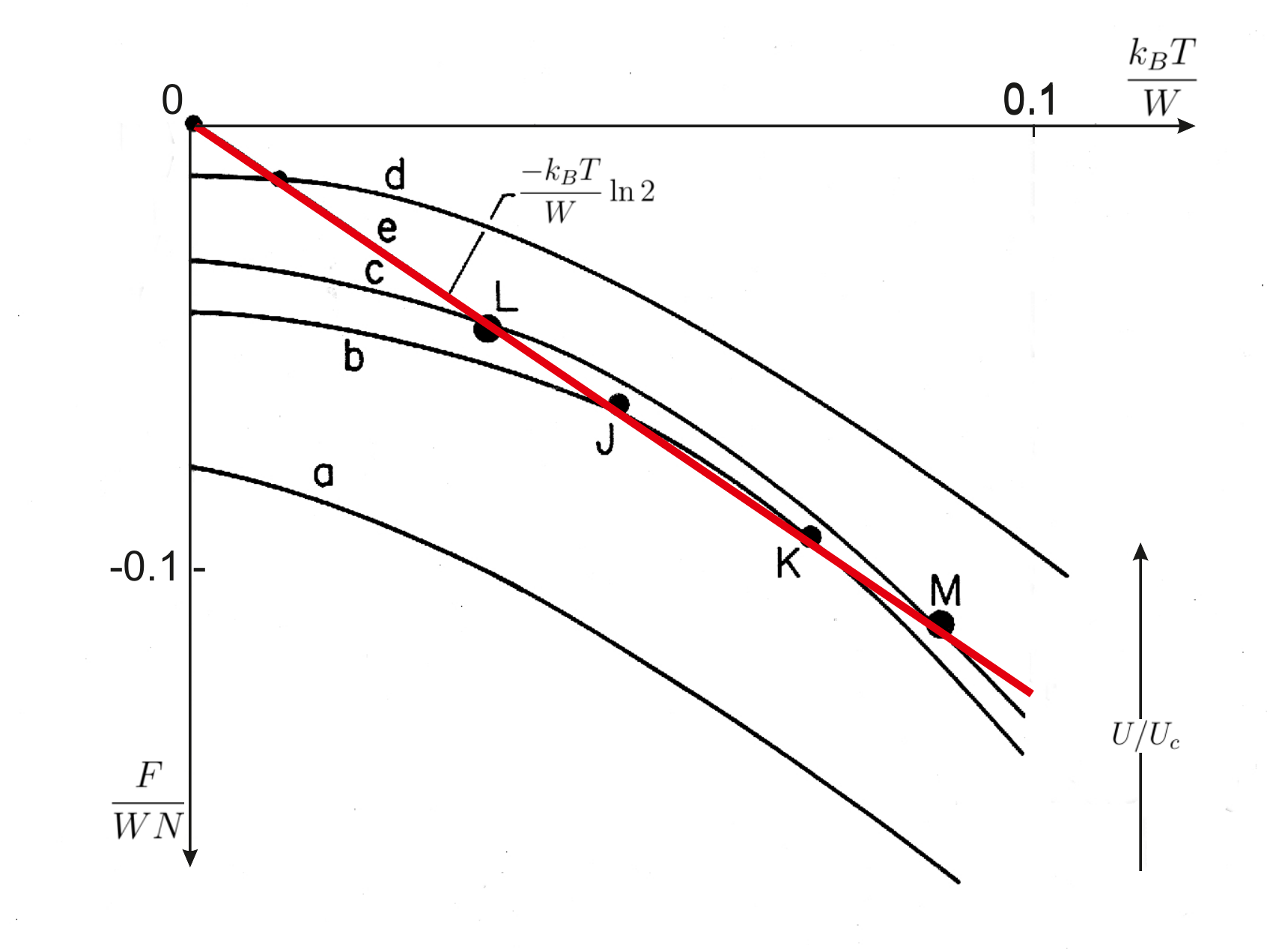}
    \caption{Temperature dependence of the free energy per particle ($F/W N$) in Fermi-liquid state (parabolas a-d), and in the Mott-Hubbard localized state (straight line e).  The crossing points LM and JK represent, respectively, M $\rightarrow$ I and I $\rightarrow$ M' transitions. In the low-temperature analysis the I $\rightarrow$ M' transition is weakly discontinuous.}
    \label{fig_2_3}
  \end{figure}
  \begin{figure}  
    \centering
   \includegraphics[width=0.6\textwidth]{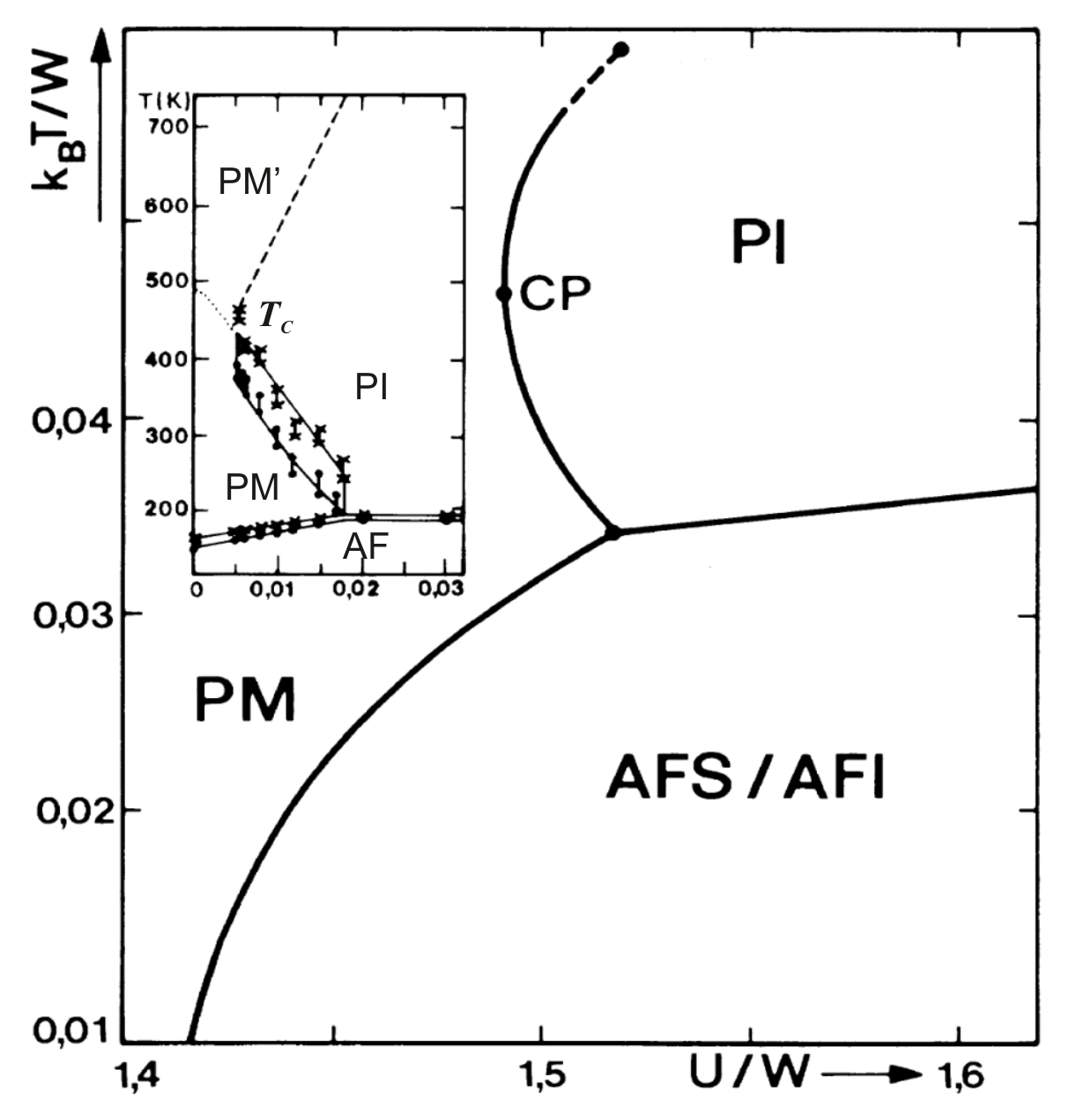}
    \caption{Phase diagram of the type presented in Fig. \ref{fig_2_2}, with inclusion of antiferromagnetic Slater (AFS) and Mott (AFI) phases. Note that $W=U_C/2$. Inset: experimentally observed \cite{KuwamotoPhysRevB1980} phase diagram on $T$-$x$ plane for \ce{(V_{1-x}Cr_{x})_{2}O_{3}}. After Ref.~\cite{SpalekPhysRevLett1987}. The quantum critical point appearing in Fig. 4a is wiped out by the presence of the antiferromagnetic order.}
    \label{Fig_6}
  \end{figure}
\noindent
The corresponding classical critical transition temperature at which the transition takes a continuous form and at $U = U_{lc}$ is
\begin{equation}
   k_B\,T_c = \frac{3\ln{2}}{2\pi^2} W\,
   \left[1-\left(\frac{U_{lc}}{U_C}\right)^2\right].
   \label{r_12+1}
\end{equation}
 For $U\leqslant U_{lc}$  the metallic (Fermi liquid) state is stable at all $T$. In effect, the regime of the transition accuracy  is determined by conditions $U_{lc} \leqslant U \leqslant U_C$.
 Disregarding the magnetic phases one then has the following overall phase sequence. For $T < T_{-}$  the system is a paramagnetic metallic (PM). For $U_{lc} < U < U_C$ and $T_{-} < T < T_{+}$ the system is a paramagnetic insulator (the lattice of fluctuating spins $S = 1/2$). For $T > T_{+}$ the re-entrant metallic behavior is observed (a crossover transition). Such a sequence is indeed observed for \ce{V_2O_3} doped with \ce{Cr} \cite{KuwamotoPhysRevB1980} and for liquid \ce{^3He} (cf. Figs 2. and 3b). The most important factor is the sequence of transformations between localized and itinerant (liquid) states of the valence electrons as a function of temperature and interaction, as shown schematically in Fig.~\ref{fig_2_2}a. For comparison, an analogical phase diagram appears for quark-gluon plasma, this time calculated as a function of the chemical potential value (cf. Fig.\ref{fig_2_2}b). 

 The physical reason for switching between the states M and I is illustrated in Fig. \ref{fig_2_3} 
 Namely, at temperature close to $T = 0$ the entropy of disordered localized moments is large ($+k_{B} \ln{2}$ per carrier), whereas for the Fermi liquid, it decreases linearly with $T$ to zero. Hence, at $T = T_{-}$ the entropy part of the free energy for localized particles outweighs that of the Fermi liquid (ALFL), even though at $T = 0$, the opposite is true. However, as the temperature is raised, the Fermi liquid entropy grows and asymptotically at high temperature approaches the value $2k_{B} \ln{2}$ per carrier. Thus, the detailed shape of the phase boundary is determined by the interplay between the competing energy and entropy contributions, as is the case for classical continuous phase transition. In summary, the continuous evolution of the system at $T = 0$ in approaching $U_C$ from below should be contrasted with the discontinuous nature of the transformation for $T>0$. Thus, the point $U = U_C$ for $T = 0$ is indeed a quantum critical point, at least within this analysis in which $d^2 =\langle n_{i\uparrow}n_{i\downarrow}  \rangle$ plays the role of the order parameter in the expression for the Ginzburg-Landau functional (\ref{r_12+1}) for almost localized correlated fermions and when the antiferromagnetic order is absent.

At the end of this Section, we would like to quote our results on metal-insulator transition including simultaneous presence of antiferromagnetism which with the increasing interaction magnitude evolves from band (\emph{Slater-type}, AFS) to the localized spin (Mott, AFI) antiferromagnetism. The part of the phase diagram depicted in Fig. \ref{fig_2_2}a 
appears only above the N\'{e}el temperature, where the antiferromagnetic states (AFS, AFI) cease to exist, here, in a discontinuous manner. The situation is shown in Fig. \ref{Fig_6}
In the inset we quote the experimental results Fig. 2b obtained for \ce{(V_{1-x}Cr_{x})_{2}O_{3}}, with the $x$ as the horizontal axis. The agreement is qualitatively good, which is rewarding since a very simple model was considered. This means that the inter-particle configurations are the crucial factor to a large extent an independent of the electronic (band) structure. 

The presence of the proposed classical critical point (CP) in Figs. \ref{fig_2_2}a 
and \ref{Fig_6} have been also beautifully confirmed much later \cite{LimeletteScience2003}. It has a mean-field character, exactly the type predicted by our mean-field-like approach \cite{SpalekPhysRevB1986,SpalekPhysRevLett1987,SpalekPhysRevB1989}, which represented the very first realistic attempt to extend theory of metal-insulator transition of the Mott-Hubbard type at $T>0$. Our results were confirmed much later \cite{GeorgesRevModPhys1996} within the dynamic mean-field approach (DMFT).

\newpage
\section{Exact diagonalization - \emph{Ab initio} approach (EDABI) to correlated systems with simple examples }

\subsection{The method}
\index{1$^{\rm{st}}$ quantization} \index{2$^{\rm{nd}}$ quantization}
The notion of a simultaneous determination of the single-particle wave-function  (1$^{\rm{st}}$ quantization aspect), combined with a precise account of inter-particle correlations (2$^{\rm{nd}}$ quantization aspect) arouse in the author's thinking about the many-particle systems because of the following circumstances. In the proper particle language in quantum mechanics (2$^{\rm{nd}}$ quantization representation) the physical particle is represented by the field operator $\hat{\Psi}_{\sigma}(\textbf{r})$ which has the form 
\begin{equation}
    \hat{\Psi}_{\sigma}(\textbf{r})\equiv \sum\Phi_{i\sigma}(\textbf{r})\,\hat{a}_{i\sigma},
\end{equation}
where the set $\{ \Phi_{i\sigma}(\textbf{r}) \}$ represents a  complete set of single particle wave-functions (not necessarily orthogonal with the corresponding set of quantum numbers $\{ i\sigma\}$ (here the spin quantum number $\sigma=\pm 1$ has been singled out explicitly to underline its fermionic nature; the whole argument holds equally well for bosons).  
The explicit 2$^{\rm{nd}}$ quantization form of the Hamiltonian is \cite{Fetter2003Quantum}
\begin{equation}
    \hat{\mathcal{H}}= \sum_{\sigma} \int d^3r\, \hat{\Psi}_{\sigma}^{\dag}(\textbf{r}) \mathcal{H}_1(\textbf{r}) 
    \hat{\Psi}_{\sigma}(\textbf{r}) + 
    \frac{1}{2} \sum_{\sigma\sigma'} \int d^3r\, d^3r'\,
    \hat{\Psi}_{\sigma}^{\dag}(\textbf{r})
    \hat{\Psi}_{\sigma'}^{\dag}(\pmb{r'})
    V(\textbf{r}-\pmb{r'})
    \hat{\Psi}_{\sigma'}^{\dag}(\pmb{r'})
    \hat{\Psi}_{\sigma}(\textbf{r}).
    \label{Hamiltonian}
\end{equation}
In this expression $\mathcal{H}_1(\textbf{r})$ represents the Hamiltonian for a single particle in wave mechanics, whereas $ V(\textbf{r}-\pmb{r'})$ is the interaction for a single pair of particles. $\hat{\Psi}_{\sigma}^{\dag}(\textbf{r})
\Psi_{\sigma}(\textbf{r})$ is the particle density operator, whereas $\hat{\Psi}_{\sigma}^{\dag}(\textbf{r})$ represents creation operator of a physical particle in the system at point $\textbf{r}$ and with spin $\sigma$. This is the reason why we call the 2$^{\rm{nd}}$ quantization representation as the particle language form of quantum approach. 

In the situation when both $\mathcal{H}_1(\textbf{r})$ and  $ V(\textbf{r}-\pmb{r'})$  are not explicitly spin independent, the Hamiltonian (\ref{Hamiltonian}) can be brought to the following form 
\begin{equation}
    \hat{\mathcal{H}}= \sum_{ij\sigma} t_{ij} \hat{a}_{i\sigma}^{\dag}\,\hat{a}_{j\sigma} +
    \frac{1}{2} \sum_{\substack{ijkl \\ \sigma\sigma'}} 
    V_{\substack{ijkl \\ \sigma\sigma'}}\, \hat{a}_{i\sigma}^{\dag}\,\hat{a}_{j\sigma'}^{\dag}\, 
    \hat{a}_{l\sigma'}\,\hat{a}_{k\sigma},
    \label{Hamiltonian_fetter-walecka}
\end{equation}
where 
\begin{equation}
    t_{ij\sigma}\equiv \langle \phi_{i\sigma} |\mathcal{H}_{1}| \phi_{j\sigma} \rangle\;\;\;  \text{and} \;\;\; 
    V_{ijkl}= \langle \phi_{i\sigma}\phi_{j\sigma'} |V|  \phi_{k\sigma'}\phi_{l\sigma} \rangle.
    \label{t_and_V}
\end{equation}

In this situation $\Phi_{i\sigma}(\textbf{r}) = \Phi_{i}(\textbf{r})\chi_{\sigma}$, we have the spin-independent hopping matrix elements $t_{ij}$, as well as the spinless interaction parameters  $V_{ijkl}$. That dynamical system behavior is determined by the corresponding operator parts of (\ref{Hamiltonian_fetter-walecka}), the matrix elements (\ref{t_and_V}) contain the arbitrary (expect complete) set of the wave functions.

In canonical modeling of the properties with the help of (\ref{Hamiltonian_fetter-walecka}) one takes into account only first few terms of the first part of $\hat{\mathcal{H}}_1(\textbf{r})$ (the hopping part) and the quantities $t_{0=t_{ii}}, t_{\langle i,j \rangle}=t$ are regarded as parameters of the model. Likewise, one takes only a few dominant terms in the interaction part and the corresponding  $V_{ijkl}$ elements  are treated also as free parameters. If one selects the Wannier basis, i.e., $\phi_{i}(\textbf{r})\equiv w_{i}(\textbf{r}) = w(\textbf{r}-\textbf{r}_{i})$,
then we can select the hoping parameters $t_{0}, t=t_{\langle i,j \rangle}$, and $t'' \equiv t_{\langle\langle i,j \rangle\rangle}$ which corresponding to the atomic 
reference energy and hoping amplitudes of particles between nearest and next nearest neighbors, respectively. 

The parametrized model created in the above way contains as a rule incomplete quantum mechanical basis, as in general the set  
$\{ w_{i}(\textbf{r})\}= \{ w_{in}(\textbf{r})\}$,
where $n$ is the type of atomic orbital and in effect the multi-orbital (multi-band) system is derived. Hence, the results may depend on the type of orbital-based model we start with. But even if this general situation is not the case, the general question is how to determine the wave functions contained in the matrix elements (\ref{t_and_V})? One way is to start from a set of orthogonalized atomic orbitals, e.g., hydrogeniclike Slater orbitals, as we discuss it below. However, the question still remains, particularly if the selected basis is not complete, whether such a basis should not be optimized in some way. This question is of crucial importance in the case of correlated systems when the two terms in (\ref{Hamiltonian_fetter-walecka}) provide contributions of the same magnitude (see the preceding Section). In particular, in the limit of strong correlations the interaction part even dominates over the single-particle contribution. 

In such a situation our proposal now is as follows. Because of the interaction predominance we first diagonalize the parametrized Hamiltonian (\ref{Hamiltonian_fetter-walecka}) in the Fock space and only subsequently minimize the obtained in such a manner ground state energy $E\equiv \langle \hat{\mathcal{H}} \rangle /N$ with respect to $i\sigma$.
We developed the whole EDABI method for the Wannier functions $\{w_{i} \}$ by treating this energy as a functional of $\{ \phi_{i\sigma} \}$.
In other words, we determine the single-particle renormalized wave-functions, now adjusted in the correlated state, by solving the effective wave equation obtained from the variational principle for the functional of the form \index{Wannier function} \index{EDABI method}
\begin{equation}
    E\{w_{i}(\textbf{r})\} \equiv E_{G} \{w_{i}(\textbf{r}) \} 
    -\mu N - \sum_{i \geqslant j} \lambda_{ij} \left( \int d^3\textbf{r}\, w^{*}_{i}(\textbf{r})\,w_{j}(\textbf{r}) -\delta_{ij}
    \right),
\end{equation}
where
\begin{equation}
N=\sum_{\sigma} \int d^3\textbf{r}\, \langle \hat{\Psi}_{\sigma}^{\dag}(\textbf{r}) \hat{\Psi}_{\sigma}(\textbf{r}) \rangle =
\sum_{ij\sigma} \int d^3\textbf{r}\, w^{*}_{i}(\textbf{r})\,w_{j}(\textbf{r}) \langle \hat{a}_{i\sigma}^{\dag} \hat{a}_{j\sigma} \rangle ,
\label{Eq_N}
\end{equation}
and $N_e=N$ is the number of particles in the system, whereas $\lambda_{ij}$ are the Lagrange multipliers, to keep the  single-particle basis orthonormal. 

The general form of this equation in the stationary case and in the ground -ensemble formalism is
\begin{equation}
\frac{\delta(E_{G}-\mu N)}{\delta w^{*}_{i}(\textbf{r})}-\nabla \frac{\delta(E_{G}-\mu N)}{\delta (\nabla w^{*}_{i}(\textbf{r}))} -
\sum_{i \geqslant j} \lambda_{ij}\,w_{j}(\textbf{r}) =0.
\end{equation}
We make a fundamental postulate concerning this equation: As this equation does not contain explicitly the (anti)commutation relations between the creation and operators, it is equally valid for both fermions and bosons and determines a rigorous, within the class of the states included in the definition of $\hat{\Psi}_{\sigma}(\textbf{r})$, the renormalized wave equation for a single-particle wavefunction in the ground state, in the milieu of remaining ($N-1$) particles.

\begin{figure} 
    \centering
   \includegraphics[width=\textwidth]{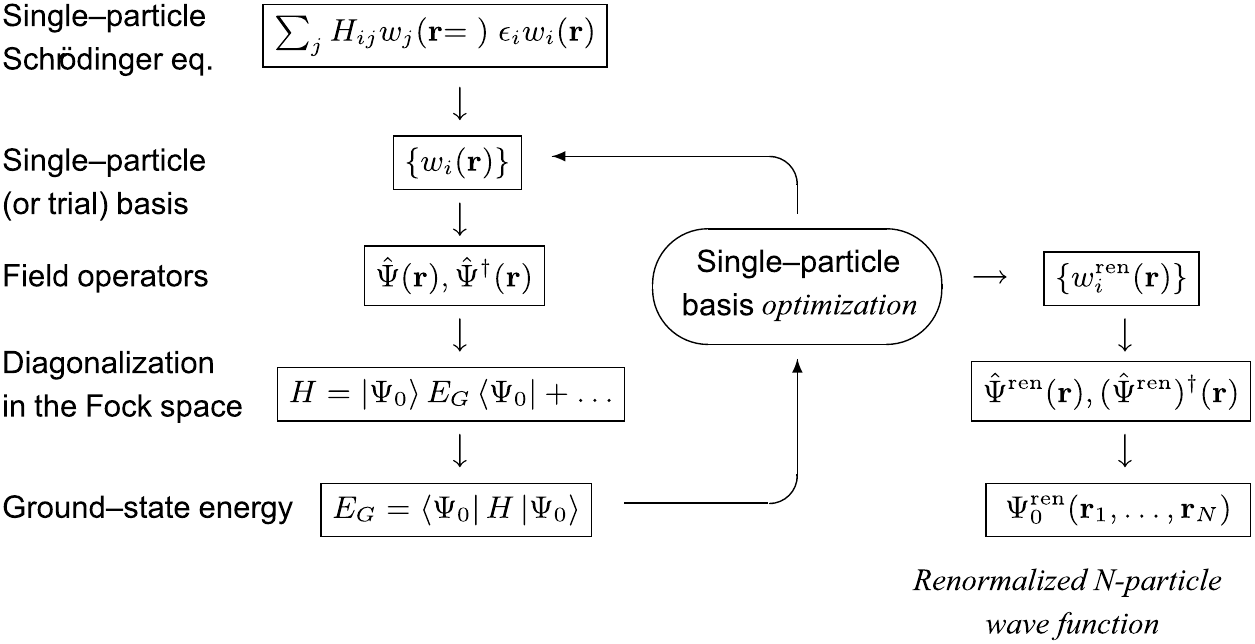}
 \caption{Flowchart describing the scheme of the EDABI method. For details see main text. When selecting the single-particle set, the topmost block should be disregarded. The renormalized many-particle wave function $\Psi_{0}^{ren}(\textbf{r}_1,\ldots, \textbf{r}_N) $ is explicitly constructed for few-particle systems in the next Section.}
\label{Fig_edabi}
\end{figure}

In this expressions the Lagrange multipliers $\lambda_{ij}$ plays the role of single-particle energy in the correlated state. Note also that the variational derivatives are taken also with respect to the averages $\langle \hat{a}_{i\sigma}^{\dag}\hat{a}_{j\sigma} \rangle$ and $\langle \hat{a}_{i\sigma}^{\dag}\hat{a}_{j\sigma'}^{\dag} \hat{a}_{l\sigma'} \hat{a}_{k\sigma} \rangle$, so 
is not just the optimization of parameters $t_{ij}$ and $V_{ijkl}$. \index{EDABI method}
Parenthetically, the same type of formal wave-function determination (together with its normalization) has been proposed by Schr\"{o}dinger in his pioneering work on wave mechanics (Schr\"{o}dinger, 1926). Also, the optimized quantity is the system energy, not the Lagrangian, which represents the classical system Hamiltonian (its expectation value)

Finally, the general $N$-particles state $|\Phi_{N}\rangle$ in the Fock space can be defined through the corresponding $N$-particle wavefunction $\Psi_{\alpha}(\textbf{r}_1\ldots, \textbf{r}_N)$ in the Hilbert space in the following manner: \cite{Robertson}
\begin{equation}
|\Phi_N \rangle= \frac{1}{\sqrt{N!}}
 \int d^3\textbf{r}_{1}\ldots \textbf{r}_{N} 
 \Psi_{N} (\textbf{r}_{1},\ldots, \textbf{r}_{N}) 
 \hat{\Psi}_{\sigma_1}^{\dag}(\textbf{r}_{1})\ldots \hat{\Psi}_{\sigma_N}^{\dag} \textbf{r}_{N}) | 0
  \rangle,
\end{equation}
where $| 0 \rangle$ is the vacuum state. One can reverse this relation and a simple algebra yields the following expression for the wavefunction $\Psi_{\alpha}(\textbf{r}_1\ldots, \textbf{r}_N)$ in the terms of $|\Phi_N \rangle$:
\begin{equation}
\Psi_{\alpha}(\textbf{r}_1\ldots, \textbf{r}_N) = \frac{1}{\sqrt{N!}} \langle 0|\hat{\Psi}_{\sigma_1}(\textbf{r}_1)\ldots \hat{\Psi}_{\sigma_N}(\textbf{r}_N)|\Phi_N \rangle,
\label{Psi_alpha}
\end{equation}
with $\alpha \equiv \{ \sigma_1,\ldots, \sigma_N   \}$. 
In other words, to obtain the wavefunction in the coordinate representation, we not only annihilate $N$ particles from the Fock state $|\Phi_N \rangle$, but also project out the thus obtained result onto the Fock vacuum state and normalize it by the factor $(N!)^{-1/2}$. Usually, the  formula (\ref{Psi_alpha}) is not used; as we proceed from the first to second quantization. Now, the crucial point is based on the observation that if we substitute in the field operator
$\hat{\Psi}(\textbf{r})$
the renormalized wavefunctions obtained from equation (\ref{Eq_N}),
then we should automatically  obtain the renormalized field operator and, as a consequence, the renormalized multiparticle wavefunction $\Psi_{\alpha}(\textbf{r}_1\ldots, \textbf{r}_N)$
from relation (\ref{Psi_alpha}). 
This last step of inserting the renormalized field operator completes the procedure of a formal treatment of the many-particle system, which avoids writing down explicitly the $N$-particle Schr\"{o}dinger equation. The approach is summarized in Fig. 7. This scheme provides an exact renormalized single-particle wavefunction from equation (\ref{Eq_N}) and the exact $N$-particle wavefunction provided we have diagonalized the second-quantized model Hamiltonian
for the problem at hand.

\subsubsection{Suplement: Finite basis approximation from the field operator: difference with the multiconfiguration interaction (MCI) approach}
\index{multiconfiguration interaction}

The field operator $\hat{\Psi}(\textbf{r})$ defined in terms of the sum over complete basis 
$\{w_i(\textbf{r})\}$ contains an infinite number of single-particle states. We assume that, in general, we represent the field operator by $M$ wavefunctions 
$\{w_i(\textbf{r})\}$. Explicitly, 
\begin{equation}
    \hat{\Psi}(\textbf{r})\equiv \sum_{i=1}^{\infty} w_i(\textbf{r})\,\hat{a}_{i} \simeq \sum_{i=1}^{M} w_i(\textbf{r})\,\hat{a}_{i},
\end{equation}
with $i$ representing a complete set of quantum numbers and $M$ being a finite number. This approximation represents one of the most fundamental features of constructing theoretical models. The neglected states usually represent highly exited (and thus negligible) states of the system. We can then write the approximate $N$-particle wavefunction ($N \leqslant M$) in the following manner 
\begin{equation}
\Psi_{\alpha}(\textbf{r}_1\ldots, \textbf{r}_n) = \frac{1}{\sqrt{N!}} 
\sum_{i_1,\ldots,i_{N}=1}^{M}
\langle 0|\hat{a}_{i_N}\ldots \hat{a}_{i_1}||\Phi_{N}\rangle w_{i_1}(\textbf{r}_1)\ldots w_{i_N}(\textbf{r}_N)
\label{Psi_alpha_r1_rN}
\end{equation}
Recognizing that within the occupation-number space spanned on the states
$\{|i_{k}\rangle \}_{k=1\ldots M}$
we have the $N$-particle state in the Fock space of the form 
\begin{equation}
|\Phi_{N}\rangle = \frac{1}{\sqrt{N!}} 
\sum_{j_1,\ldots,j_{N}=1}^{M}
C_{j_{1}\ldots j_{N}}
\hat{a}_{j_1}^{\dag}\ldots \hat{a}_{j_N}^{\dag}|0\rangle,
\label{Phi}
\end{equation}
where $C_{j_{1}\ldots j_{N}}$ represents  the the coefficients of the expansion to be determined from a diagonalization procedure. Substituting (\ref{Phi})  into (\ref{Psi_alpha_r1_rN}) we obtain
\begin{equation}
\Psi_{\alpha}(\textbf{r}_1\ldots, \textbf{r}_n) = \frac{1}{\sqrt{N!}} 
\sum_{i_1,\ldots,i_{N}=1}^{M}
\sum_{j_1,\ldots,j_{N}=1}^{M}
\langle 0| \hat{a}_{i_1}\ldots \hat{a}_{i_N}
\hat{a}_{j_1}^{\dag}\ldots \hat{a}_{j_N}^{\dag}|0\rangle 
C_{j_{1}\ldots j_{N}}
 w_{i_1}(\textbf{r}_1)\ldots w_{i_N}(\textbf{r}_N).
\end{equation}

The expression provides $N!$ nonzero terms each equal to $(-1)^{P}$, where $P$ represents the sign of the permutation of quantum numbers ($j_{1}\ldots j_{N}$) with respect to a selected collection ($i_{1}\ldots i_{N}$). In other words, we can write that
\begin{equation}
\Psi_{\alpha}(\textbf{r}_1\ldots, \textbf{r}_n) = \frac{1}{\sqrt{N!}} 
\sum_{i_1,\ldots,i_{N}=1}^{M}
C_{i_{1}\ldots i_{N}}(A,S)
 w_{i_1}(\textbf{r}_1)\ldots w_{i_N}(\textbf{r}_N).
\end{equation}
We have the same expansion coefficients for both the wavefunction in the Fock space $| Phi_N\rangle$ and that in the Hilbert space $\Psi_{\alpha}(\textbf{r}_1\ldots, \textbf{r}_n)$! 
Therefore, the above expression represents the multiconfigurational-interaction wavefunction of $N$ particles distribution among M states with the corresponding weights $C_{i_{1}\ldots i_{N}}$
for each configuration and ($A,S$) represents respectively the antisymmetrization (Slater determinant) or the symmetrization (simple product 
$ w_{i_1}(\textbf{r}_1)\ldots w_{i_N}(\textbf{r}_N)$
for the fermions and bosons, respectively. Whereas the MCI used in quantum chemistry [....] is based on variational optimizations of the coefficients $C_{i_{1}\ldots i_{N}}$ and
, here the coefficients $C$ are determined from diagonalization in the Fock space, spanned on $M$ states in the Hilbert space. The presence of wave equation (\ref{Eq_N})
thus supplements the usual MCI approach. 

Next, we discuss first selected elementary examples from atomic and molecular physics, before turning to modeling extended correlated systems.

\subsection{Elementary examples from atomic physics}
 
One of the principal attractive features of EDABI method is the ability to construct atomic orbitals with a precise account for inter-electronic interactions. Here this program is illustrated on example of lightest atomic systems, as well as by an elementary example of the wave equation for renormalized wave functions.

\subsubsection{A didactic example: \ce{He} and \ce{Li} atom} \index{EDABI method}

We start by selecting as $\{ w_i({\bf r})\}$ just two $1s$-type Slater orbitals for the He atom $\Phi_\sigma({\bf r}) = (\alpha^3/\pi)^{1/2} \exp(-\alpha r) \chi_\sigma$, where $\alpha$ is the effective inverse radius of the states. In other words, the simplest trial field operator is of the form
\begin{equation}
\hat{\Phi}({\bf r}) = \Phi_\uparrow ({\bf r}) \hat{a}_\uparrow +
\Phi_\downarrow({\bf r}) \hat{a}_\downarrow , \label{Psi1s}
\end{equation}
where $a_\sigma$ is the annihilation operator of particle in the
state $\Phi_\sigma({\bf r})$. The Hamiltonian in the second
quantization for this two-element basis has then the form
\begin{equation}
H = \epsilon_a ( \hat{n}_\uparrow +\hat{n}_\downarrow ) + U \hat{n}_\uparrow \hat{n}_\downarrow ,
\label{HHe1s}
\end{equation}
where $\hat{n}_\sigma = \hat{a}^\dagger_\sigma \hat{a}_\sigma$, whereas
\begin{equation}
\epsilon_a = \langle \Phi_\sigma \vert H_1  \vert \Phi_\sigma
\rangle ,
\end{equation}
and
\begin{equation}
U = \langle \Phi_\sigma \Phi_{\overline{\sigma}} \vert V  \vert
\Phi_{\overline{\sigma}} \Phi_\sigma \rangle
\end{equation}
are the matrix elements of the single-particle part defined as
\begin{equation}
H_1 = -\frac{\hbar^2}{2 m} \nabla_1^2 -\frac{\hbar^2}{2 m}
\nabla_2^2 - \frac{2 e^2}{\kappa_0 r_1} - \frac{2 e^2}{\kappa_0
r_2} \stackrel{a.u.}{\equiv} -\nabla_1^2 - \nabla_2^2 -
\frac{4}{r_1} - \frac{4}{r_2}
\end{equation}
and of the Coulomb interaction
\begin{equation}
V = \frac{e^2}{\kappa_0 |{\bf r_1} - {\bf r_2}|}
\stackrel{a.u.}{\equiv} \frac{2}{|{\bf r_1} - {\bf r_2}|} ,
\end{equation}
with the corresponding definitions in atomic units after the second equality sign. The only eigenvalue of (\ref{HHe1s}) is obtained for the state $\hat{a}^\dagger_\uparrow \hat{a}^\dagger_\downarrow |0>$ and is $E = 2 \epsilon_a + U$. This total energy is then minimized with respect
to $\alpha$ to obtain the well-known Bethe and Salpeter variational estimate of both $\alpha$ and the ground state energy $E_G$, as discussed before. However, we may look at the problem
differently. 
The true wavefunction is obtained from the Euler equation for the functional $E=E\{\Psi_{\sigma}\}$ under the proviso that the wave function is normalized. This means that we minimize the functional
\begin{equation}
E\{ \Phi_\sigma({\bf r}) \} = \sum_\sigma \int d^3r
\Phi^*_\sigma({\bf r}) H_1({\bf r}) \Phi_\sigma({\bf r})
+ \frac{1}{2} \sum_\sigma \int d^3r d^3r' |\Phi_\sigma({\bf r})|^2
V_{12}({\bf r} - {\bf r'}) |\Phi_{\overline{\sigma}}({\bf r})|^2.
\end{equation}
In effect, renormalized wave equation take the form of the unrestricted Hartree equations for $\Phi_\sigma ({\bf r})$
\begin{equation}
\left( \nabla^2 - \frac{2 e^2}{\kappa_0 r} \right)\Phi_\sigma({\bf
r})+ \Phi_\sigma({\bf r}) \int d^2r' \frac{e^2}{\kappa_0 |{\bf r}
-{\bf r'}|} |\Phi_{\overline{\sigma}}({\bf r'})|^2 = \lambda
\Phi_\sigma({\bf r}) . \label{HFeqn}
\end{equation}
Thus, we can see that taking in the simplest case just two spin $1s$-type
orbitals we obtain either the well-known Bethe-Salpeter variational estimate 
for $\alpha$ and $E_G$  for He atom: $\alpha = 27/(16 a_0)$ and $E_G = -5.695 Ry$, where $a_0 \simeq 0.53$\AA \ is the $1s$ Bohr orbit radius. We see that the He
atom is the smallest in the Universe!\\
\indent The proposed expression (\ref{Psi1s}) for the
field operator is the simplest one, but it leads to nontrivial results even though  the trial atomic basis $\{ \Phi_\sigma({\bf r}) \}$ is far from being complete in the quantum-mechanical sense. However, we can improve systematically on the basis by selecting a richer basis than that in (\ref{Psi1s}). The further step in this direction is discussed next.
Namely, we can expand the field operator in the basis involving the higher order irreducible representations of the rotation group with $n=2$, which in the variational scheme involve including, apart from the $\Psi_{1s}({\bf r})$ orbital, also the higher $\Psi_{2s}({\bf r})$ and $\Psi_{2pm}({\bf r})$ orbitals, with $m = \pm 1,0$ (i.e., the next shell); all of them involving the adjustment of the corresponding orbital characteristics $\alpha_i$, $i = 1s, 2s$ and $2pm$. The field operator is then
\begin{equation}
\hat{\Psi}({\bf r}) = \sum_\sigma \left[ w_{1s} ({\bf r}) \chi_{1
\sigma} \hat{a}_{1 \sigma} + w_{2s}({\bf r}) \chi_{2 \sigma} \hat{a}_{2 \sigma}
+ \sum_{m = -1}^{+1} w_{2pm}({\bf r}) \chi_{m \sigma} \hat{a}_{2pm \sigma}
\right]
\equiv \sum_{i \sigma} w_i({\bf r}) \chi_{i
\sigma} \hat{a}_{i \sigma} , \label{Psi1s2s2pm}
\end{equation}
where $w_i({\bf r})$ are orthogonalized orbitals obtained from the nonorthogonal atomic \footnote{Note that the atomic orbitals $1s$ and \it{2s} are not orthogonal to each other for arbitrary values of their spatial extents $1/\alpha_i$. The $2p$ orbitals are orthogonal to each other and to s orbitals,
since they contain a nontrivial angular dependence expressed via
spherical harmonics $Y^m_l (\theta,\phi)$.} basis $\{ \Psi_i({\bf r}) \}$ in a standard manner. The Fock space spanned on $2 + 2 + 6 = 10$ trial spin orbitals contains $D = { 2 M \choose N_e }$ dimensions, where $M = 5$ now and $N=N_e = 2, 3$ is the number of electrons for \ce{He} and \ce{Li}, respectively. This means that $D = 45$ and $120$ in
those two cases and we have to diagonalize the Hamiltonian
matrices of that size to determine the ground and the lowest
excited states.

\begin{table}
\caption{Optimized Bohr-orbit radii $a_i = \alpha_i^{-1}$ of $1s$, \it{2s},
and $2p$ orbits (in units of $a_0$), the overlap $S$ between
renormalized $1s$ and \it{2s} states, and the ground state energy for the
lightest atoms and ions (five Slater orbitals taken).}
\label{Toptradii} \begin{center}
\begin{tabular}{cccccc}
\hline \hline
 & $a_{1s}$ & $a_{2s}$ & $a_{2p}$ & $S$ & $E_G$ (Ry) \\
\hline $H$ & 1 & 2 & 2 & 0 & -1 \\
 $H^-$ & 0.9696 & 1.6485 & 1.017 & -0.1 & -1.0487 \\
 $He$  & 0.4274 & 0.5731 & 0.4068 & -0.272 & -5.79404 \\
 $He^-$ & 1.831 & 1.1416 & 0.4354 & -0.781 & -5.10058 \\
 $Li$ & 0.3725 & 1.066 & 0.2521 & 0.15 & -14.8334 \\
 $Be^+$ & 0.2708 & 0.683 & 0.1829 & 0.109 & -28.5286 \\
\hline \hline
\end{tabular}
\end{center}
\end{table}
One should note that we construct and subsequently diagonalize the $\langle i|\hat{\mathcal{H}}|j\rangle$ matrix in the Fock space for (fixed) parameters $\epsilon_a, t_{ij}$, and $V_{kl}$. After the diagonalization has been carried out, we readjust the wave function and start the
whole procedure again until the absolute minimum is reached (cf. Fig. \ref{Fig_edabi}).

By diagonalizing the corresponding Hamiltonian matrices and
subsequently, minimizing the lowest eigenvalue with respect to the parameters $\alpha_i$ - the inverse radial extensions of the
corresponding wave functions, we obtain the results presented in
Table \ref{Toptradii} (the values $a_{2pm}$ are all equal within the numerical accuracy $\sim10^{-6})$. For example, the ground state energy of He is $E_G = -5.794$ Ry, which is close to the accepted "exact" value 
$-5.8074$, given  the simplicity of our approach. Further improvement is feasible by either including the $n=3$ states or by resorting to the Gaussian trial basis. These are not analyzed here. 

First, we can represent the ground-state two-particle
spin-singlet wavefunction for He atom taking $\hat{\Psi}({\bf r|})$ in the form (\ref{Psi1s2s2pm}), which has the following form \cite{GorlichPhDThesis}
\begin{eqnarray}
|\Psi^{He}_0\rangle \simeq ( -0.799211 \hat{a}^+_{1s\downarrow} \hat{a}^+_{1s\uparrow} + 0.411751 \hat{a}^+_{1s\downarrow} \hat{a}^+_{2s\uparrow} - 0.411751 \hat{a}^+_{1s\uparrow} \hat{a}^+_{2s\downarrow}
\\ \nonumber
- 0.135451a^+_{2s\downarrow} \hat{a}^+_{2s\uparrow} + 0.0357708
\hat{a}^+_{2p0\downarrow} \hat{a}^+_{2p0\uparrow} + 0.0357641
\hat{a}^+_{2p1\downarrow} \hat{a}^+_{2p-1\uparrow}
\\ \nonumber
- 0.0357641 \hat{a}^+_{2p1\uparrow}
\hat{a}^+_{2p-1\downarrow} ) |0\rangle,
\end{eqnarray}
Similarly, the $S^z = + 1/2$ state for Li atom is of the form
\begin{eqnarray}
|\Psi^{Li}_0\rangle \simeq ( 0.997499 \hat{a}^+_{1s\downarrow} \hat{a}^+_{1s\uparrow}
\hat{a}^+_{2s\uparrow} -0.0570249 \hat{a}^+_{1s\uparrow} \hat{a}^+_{2s\downarrow}
\hat{a}^+_{2s\uparrow}
\\ \nonumber
+ 0.0039591 \hat{a}^+_{1s\uparrow} \hat{a}^+_{2p0\downarrow} \hat{a}^+_{2p0\uparrow} +
0.00395902 \hat{a}^+_{1s\uparrow} \hat{a}^+_{2p1\downarrow} \hat{a}^+_{2p-1\uparrow}
\\ \nonumber
-0.00395894 \hat{a}^+_{1s\uparrow} \hat{a}^+_{2p1\uparrow} \hat{a}^+_{2p-1\downarrow}
- 0.023783 \hat{a}^+_{2s\uparrow} \hat{a}^+_{2p0\downarrow} \hat{a}^+_{2p0\uparrow}
\\ \nonumber
-0.0237806 \hat{a}^+_{2s\uparrow} \hat{a}^+_{2p1\downarrow} \hat{a}^+_{2p-1\uparrow}
+0.0237806 \hat{a}^+_{2s\uparrow} \hat{a}^+_{2p1\uparrow} \hat{a}^+_{2p-1\downarrow}
) |0\rangle .
\end{eqnarray}
We see that the probability of encountering the configuration $1s^2$ in He is less than $2/3$, whereas the corresponding configuration $1s^2 2s$ for \ce{Li} almost coincides with that for the hydrogenic-like picture. The reason for the difference is that the overlap integral between $1s$ and $2s$ states $S = \langle 1s|2s \rangle$ in the former case is
large and the virtual transitions $1s \rightleftharpoons 2s$ do not involve a substantial change in of the Coulomb energy. Those wave
functions can be used to evaluate any ground-state characteristic by
calculating $<\Psi_G|\hat{O}|\Psi_G>$ for $\hat{O}$ represented in
the 2$^{nd}$ quantized form. For example, the atom dipole moment operator
is $\hat{\bf d} = e \int d^3r
\hat{\Psi}^\dagger({\bf r}) {\bf x} \hat{\Psi}({\bf r})$, etc.\\
\indent The second feature is connected with determination of the
microscopic parameters $V_{ijkl}$ in our Hamiltonian, since their
knowledge is crucial for atomic cluster calculations, as well as
the determination of physical properties of extended systems as a
function of the lattice parameter.
 Namely, we
can rewrite the Hamiltonian 
for the case of single
atom within the basis (\ref{Psi1s2s2pm}) in the form
\begin{equation}
\begin{split}
\mathcal{H} = \sum_{i \sigma} \epsilon_i \hat{n}_{i \sigma} + t \sum_{\sigma} \left( \hat{a}^\dagger_{2 \sigma} \hat{a}_{1 \sigma} + \hat{a}^\dagger_{1 \sigma}
\hat{a}_{2 \sigma} \right) + \sum_{i = 1}^{5} U_i
   \hat{n}_{i \uparrow} \hat{n}_{i \downarrow} +
\frac{1}{2} \sum_{i \neq j} K_{i j} \hat{n}_i \hat{n}_j
\\
 -\frac{1}{2}\sum_{ i
\neq j} J_{i j} \left( {\bf S_i \cdot S_j} - \frac{1}{2} \hat{n}_i \hat{n}_j
\right) + \sum_{i \neq j} J_{ij} \hat{a}^\dagger_{i \uparrow} \hat{a}^\dagger_{i \downarrow} \hat{a}_{j \downarrow} \hat{a}_{j \uparrow}
+ \sum_{i \neq j \sigma} V_{ij} \hat{n}_{i \overline{\sigma}} \hat{a}^\dagger_{i \sigma}  \hat{a}_{j \sigma}.
\label{ParHam}
\end{split}
\end{equation}
$t$ is the hopping integral between $1s$ and $2s$ states, $U_i$ are the intraorbital Coulomb interactions, $K_{i j}$ are their
interorbital correspondents, $V_{ij}$ is the so-called correlated hopping integral, and $J_{ij}$ is the direct exchange integral, for states $i$ and $j = 1,\ldots,5$. The principal parameters for the atoms and selected ions are provided in Table \ref{Tmicropar}. We can draw the following interpretation from this analysis. The calculated energy difference $\Delta E$ for \ce{He} between the ground-state singlet and the first excited triplet is $-2.3707 - (-5.794) \simeq 3.423$Ry (the singlet $1s\uparrow 2s\downarrow$
is still 1 Ry higher). The corresponding energy of the Coulomb
interaction in the $1s^2$ configuration is $U_1 = 3.278Ry$, a value comparable to $\Delta E$. Additionally, the Coulomb interaction in $1s\uparrow 2s\downarrow$ state is $\approx 1.5 Ry$, a substantially lower value. The relative energetics tells us why we have a substantial admixture of the excited $1s\uparrow 2s\downarrow$ state to the singlet $1s^2$. In other words, a substantial Coulomb interaction ruins hydrogenic-like scheme, although the actual values could be improved further by enriching the trial basis.

\begin{table}
\caption{Microscopic parameters (in Ry) of the selected atoms and ions all quantities are calculated for the orthogonalized atomic states. $t$ is the $1s-2s$ hopping magnitude, $U_i$ is the intraorbital Coulomb interaction ($i=1s(1), 2s(2), m=0(3)$, and $m=\pm1(p)$), whereas $K_{ij}$ and $J_{ij}$ are the interorbital Coulomb and exchange interaction parameters. }
\label{Tmicropar}
\begin{tabular}{ c  c  c c c c  c c c c c c}
\hline \hline
 & $t$ & $U_1$ & $U_2$ & $U_3$ & $U_p$ & $K_{12}$ & $K_{13}$ & $K_{23}$ & $J_{12}$ & $J_{13}$ & $J_{23}$  \\
\hline $H^-$ & 0.057 & 1.333 & 0.369 & 0.77 & 0.728 & 0.519 & 0.878 & 0.457 & 0.061 & 0.138 & 0.035 \\
 $He$ & 1.186 & 3.278 & 1.086 & 1.924 & 1.821 & 1.527 & 2.192 & 1.289 & 0.212 & 0.348 & 0.115 \\
 $He^-$ & -1.1414 & 1.232 & 0.764 & 1.798 & 1.701 & 0.929 & 1.421 & 1.041 & 0.269 & 0.28 & 0.102 \\
 $Li$ & -0.654 & 3.267 & 0.533 & 3.105 & 2.938 & 0.749 & 3.021 & 0.743 & 0.06 & 0.606 & 0.014 \\
 $Be^+$ & -0.929 & 4.509 & 0.869 & 4.279 & 4.049 & 1.191 & 4.168 & 1.175 & 0.105 & 0.837 & 0.025 \\
\hline \hline
\end{tabular}
\end{table}

One may ask how the renormalized wave equation would look
in the present situation. The answer to this question is already not brief for the basis containing $M=5$ starting states $\{w_i({\bf r}) \}$ and will not be tackled here.

\begin{figure}
    \centering
   \includegraphics[width=0.6\textwidth]{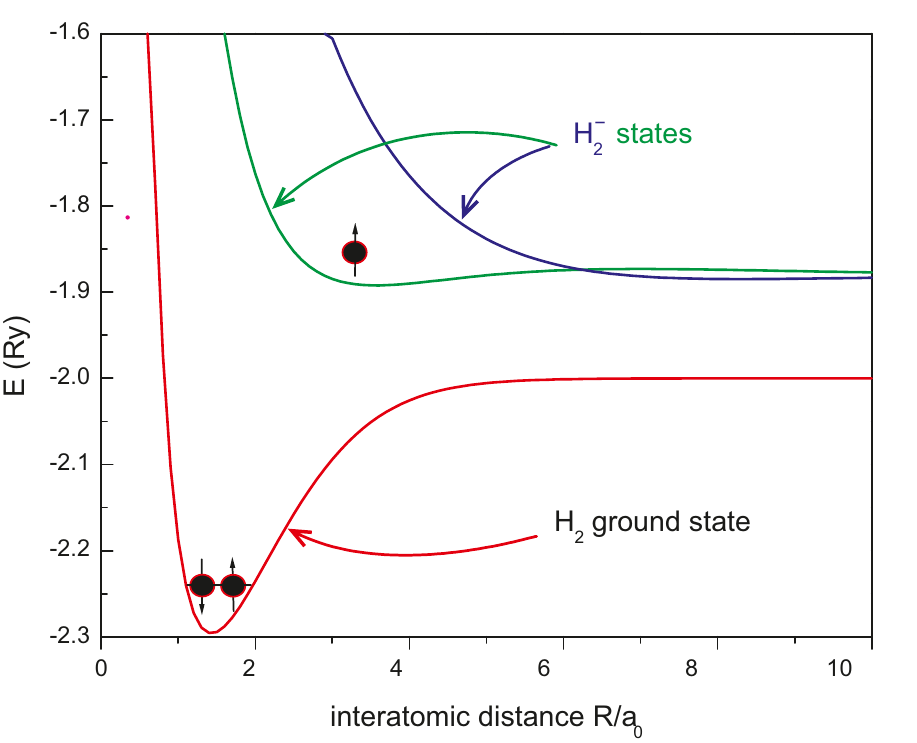}
 \caption{The level scheme of the \ce{H_2} ground state and the lowest \ce{H_2^-} states as a function of the interatomic distance $R$.} \label{Fig_8}
\end{figure}

\subsection{\ce{H_2} molecule and \ce{H_2^-} ion}

In this Subsection we consider \ce{H_2} molecule. For the illustration of the method we have plotted in Fig. \ref{Fig_8} the level scheme for the \ce{H_2} and \ce{H_2^-} systems. We consider first the situation with only one
$1s$-like orbital per atom. For \ce{H_2} we have $ { 4 \choose 2 } = 6$ two particle states. For that purpose, we start with the parameterized Hamiltonian (\ref{ParHam}), where subscripts $'i'$ and $'j'$ label now the two atomic sites and hence $U_1 = U_2 = U$, $ K_{12} = K$, $J_{12} = J$, $V_{12} = V$, and
$\epsilon_1 = \epsilon_2 = \epsilon_a$. Note that the Hamiltonian  (\ref{ParHam}) in the two -site (\ce{H_2}) case contains all possible intersite interactions.

\begin{figure} 
\resizebox{0.85\columnwidth}{!}{%
  \includegraphics{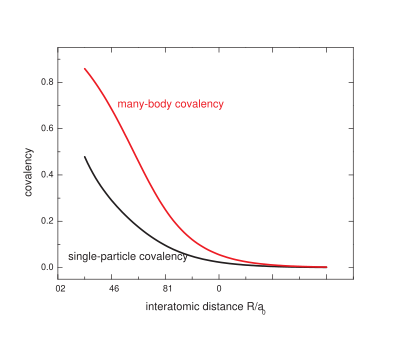}
} \caption{The single-particle ($\gamma$) and many-body
($\gamma_{mb}$) covalency factors for the \ce{H_2} wave functions. For details see main text. Note that the many-body covalency is stronger then its single-particle corespondant (orbital mixing). } \label{Fig3}
\end{figure}

The lowest eigenstate for \ce{H_2} is the spin-singlet state
\begin{equation}
E_G \equiv \lambda_5 = 2 \epsilon_a + \frac{1}{2} ( U + K ) + J -
\frac{1}{2} \left[ ( U - K )^2 + 16 ( t + V )^2 \right]^{1/2} ,
\label{lambda5}
\end{equation}
and the corresponding singlet ground state in the Fock space has the form
\begin{equation}
\begin{split}
| G \rangle = \frac{1}{ \sqrt{2 D (D - U + K)} } \times
\\
\left\{  \frac{4 ( t + V )}{ \sqrt{2} } ( \hat{a}^\dagger_{1 \uparrow}
\hat{a}^\dagger_{2 \downarrow}- \hat{a}^\dagger_{1 \downarrow} \hat{a}^\dagger_{2
\uparrow} ) -  \frac{( D - U + K)}{\sqrt{2}}( \hat{a}^\dagger_{1 \uparrow}
\hat{a}^\dagger_{2 \downarrow} + \hat{a}^\dagger_{1 \downarrow} \hat{a}^\dagger_{2
\uparrow} ) \right\} | 0 \rangle , \label{Gvect}
\end{split}
\end{equation}
where 
\[
D \equiv \left[ (U - K)^2 + 16 (t + V)^2 \right] ^ {1/2}.
\]
The lowest spin-singlet eigenstate has an admixture of the ionic state $\frac{1}{\sqrt{2}}( \hat{a}^\dagger_{1 \uparrow} \hat{a}^\dagger_{2
\downarrow} + \hat{a}^\dagger_{1 \downarrow} \hat{a}^\dagger_{2 \uparrow} )$. Therefore, to see the difference with either the Hartree-Fock or Heitler-London approach to \ce{H_2} is that we construct the two-particle
wavefunction for the ground state according to the prescription
\begin{equation}
\Phi_0 ({\bf r}_1,{\bf r}_2) \equiv \frac{1}{\sqrt{2}} \langle 0| \hat{\Psi} ({\bf r}_1) \hat \Psi ({\bf r}_2) | G \rangle.
\label{GSprescript}
\end{equation}
Taking $\hat \Psi ({\bf r}) = \sum_{i = 1}^2 \sum_{\sigma =
\uparrow}^\downarrow \Phi_i({\bf r}) \chi_\sigma ({\bf r})$, we
obtain that
\begin{equation}
\Phi_0 ({\bf r}_1,{\bf r}_2) = \frac{2 ( t + V )}{\sqrt{2 D (D - U
+ K)}} \Phi_c ({\bf r}_1,{\bf r}_2) - \frac{1}{2} \sqrt{\frac{D - U
+ K}{2 D}} \Phi_i ({\bf r}_1,{\bf r}_2),
\end{equation}
where the covalent part is
\begin{equation}
\Phi_c ({\bf r}_1,{\bf r}_2) =
 \left[ w_1({\bf r}_1) w_2({\bf
r}_2) + w_1({\bf r}_2) w_2({\bf
r}_1) \right] \left[ \chi_\uparrow({\bf r}_1) \chi_\downarrow({\bf r}_2) - \chi_\downarrow({\bf r}_1) \chi_\uparrow({\bf
r}_2) \right],
\end{equation}
whereas the ionic part takes the form
\begin{equation}
\Phi_i ({\bf r}_1,{\bf r}_2) =
\left[ w_1({\bf r}_1) w_1({\bf
r}_2) + w_2({\bf r}_1) w_2({\bf r}_2) \right] \left[
\chi_\uparrow({\bf r}_1) \chi_\downarrow({\bf r}_2) -
\chi_\downarrow({\bf r}_1) \chi_\uparrow({\bf r}_2) \right].
\end{equation}
The ratio of the coefficients before $\Phi_c ({\bf r}_1,{\bf r}_2)$ and $\Phi_i ({\bf r}_1,{\bf r}_2)$ can be termed as the \emph{many-body covalency} $\gamma_{mb}$. This value should be distinguished from the usual \emph{single-particle covalency} $\gamma$
appearing in the definition of the orthogonalized atomic orbital $w_i({\bf r})$:
\begin{equation}
w_i({\bf r}) = \beta \left[ \Phi_i({\bf r}) - \gamma \Phi_j({\bf
r})\right],
\end{equation}
with $j \neq i$. The two quantities are drawn in Fig. \ref{Fig3}. The many-body covalency $\gamma_{mb}$ represents a true degree of multiparticle configurational mixing.

In Table \ref{TH2micro} we list the energies and the values of the microscopic parameters for \ce{H_2} system with optimized orbitals. One should notice a drastic difference for the so-called {\em correlated hopping} matrix element $V$ in the two cases. The same holds true for the direct exchange integral $J$
(ferromagnetic). This exchange integral is always decisively smaller than that for the antiferromagnetic kinetic exchange, $J_{kex} \equiv 4
(t+V)^2/(U - K)$. 

\begin{table}
\caption{Ground-state energy and microscopic parameters (in Ry) for \ce{H_2} molecule. The last column represents the kinetic exchange integral characterizing intersite antiferromagnetic exchange}
\label{TH2micro}
\begin{tabular}{rcccccccc} \hline \hline $R/a$ & $E_G/N$ & $\epsilon_a$ & $t$ & $U$ & $K$ & $V$ [mRy] & $J$ [mRy] &
$\frac{4(t+V)^2}{U-K}$[mRy]
 \\  \hline 1.0 & -1.0937 & -1.6555 & -1.1719 & 1.8582 & 1.1334 & -13.5502 &
   26.2545 & 7755.52 \cr 1.5 & -1.1472 & -1.7528 & -0.6784 & 1.6265 & 0.9331 &
    -11.6875 & 21.2529 & 2747.41 \cr 2.0 & -1.1177 & -1.722 & -0.4274 & 1.4747 &
   0.7925 & -11.5774 & 16.9218 & 1130.19 \cr 2.5 & -1.0787 & -1.6598 & -0.2833 &
   1.3769 & 0.6887 & -12.0544 & 13.1498 & 507.209 \cr 3.0 & -1.0469 & -1.5947 &
    -0.1932 & 1.3171 & 0.6077 & -12.594 & 9.8153 & 238.939 \cr 3.5 & -1.0254 &
    -1.5347 & -0.1333 & 1.2835 & 0.5414 & -12.8122 & 6.9224 & 115.143 \cr 4.0 &
    -1.0127 & -1.4816 & -0.0919 & 1.2663 & 0.4854 & -12.441 & 4.5736 &
   55.8193 \cr 4.5 & -1.006 & -1.4355 & -0.0629 & 1.2579 & 0.4377 & -11.4414 &
   2.8367 & 26.9722 \cr 5.0 & -1.0028 & -1.3957 & -0.0426 & 1.2539 & 0.3970 &
    -9.9894 & 1.6652 & 12.9352 \cr 5.5 & -1.0012 & -1.3616 & -0.0286 & 1.2519 &
   0.3623 & -8.3378 & 0.9334 & 6.1455 \cr 6.0 & -1.0005 & -1.3324 &
    -0.01905 & 1.251 & 0.3327 & -6.7029 & 0.5033 & 2.8902 \cr 6.5 & -1.00024 &
    -1.3073 & -0.0126 & 1.2505 & 0.3075 & -5.2242 & 0.2626 & 1.3452 \cr 7.0 &
    -1.0001 & -1.2855 & -0.0083 & 1.2503 & 0.2856 & -3.9685 & 0.1333 &
   0.6197 \cr 7.5 & -1.00004 & -1.2666 & -0.0054 & 1.2501 & 0.2666 &
    -2.9509 & 0.066 & 0.2826 \cr 8.0 & -1.00002 & -1.25 & -0.0035 &
   1.25006 & 0.25 & -2.1551 & 0.032 & 0.1277 \cr 8.5 & -1.00001 & -1.2353&
    -0.0023 & 1.25003 & 0.2353 & -1.5501 & 0.01523 & 0.0572 \cr 9.0 & -1. &
    -1.2222 & -0.0015 & 1.25001 & 0.2222 & -1.1005 & 0.0071 &
   0.0254 \cr 9.5 & -1. & -1.2105 & -0.0009 & 1.25001 & 0.2105 & -0.7725 &
   0.0033 & 0.0112 \cr 10.0 & -1. & -1.2 & -0.0006 & 1.25 & 0.2 & -0.5371 &
   0.0015 & 0.0049 \cr   \hline \hline
\end{tabular}
\end{table}

\subsubsection{Hydrogen clusters \ce{H_N}}
\begin{figure}          
    \centering
   \includegraphics[width=0.9\textwidth]{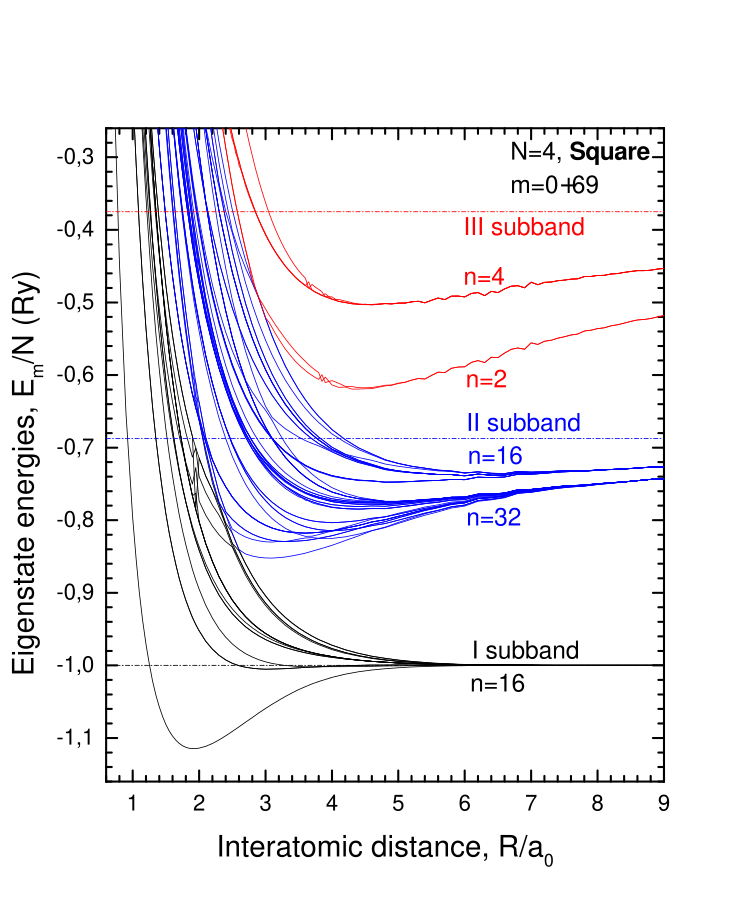}
\caption{Ground- and excited-states energies for the \ce{H_4} square configuration as a function of the interatomic distance. The position of subsequent Hubbard subbands (distant by $U$) are marked. The lowest two are homo-lumo split energy levels. 
}
\label{Fig4}
\end{figure}

\begin{figure} 
    \centering
   \includegraphics[width=0.9\textwidth]{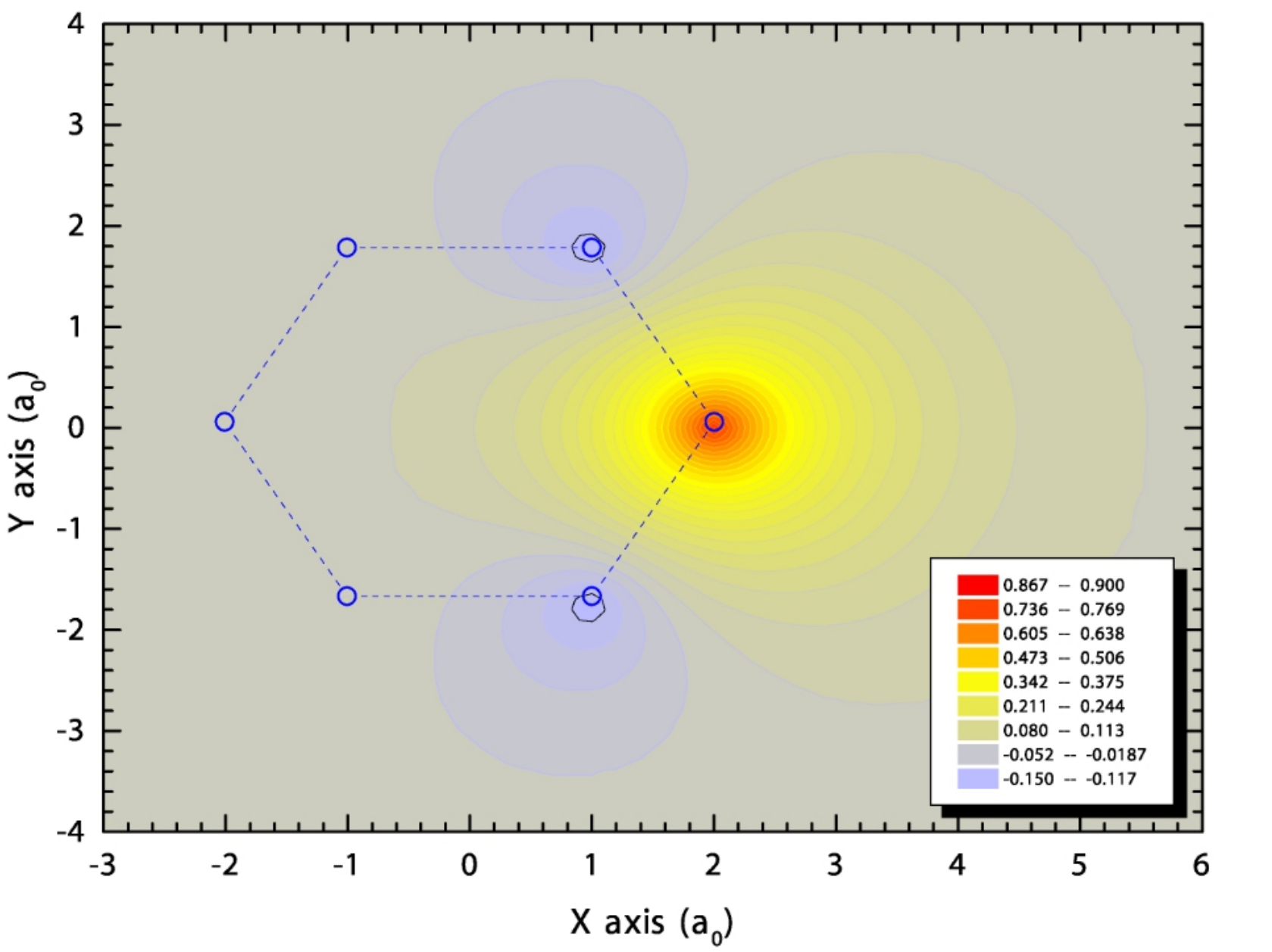}.
    \caption{Single Wannier function for the \ce{H_6} configuration for the ground state and at the optimal interatomic distance. Note the negative values at the nearest neighboring sites, as well as its anisotropic character due to the system geometric 6-fold symmetry.}
    \label{Fig_Phd_Zach}
  \end{figure}

As the next application we consider hydrogen-cluster \ce{H_N} systems. We take the atomic-like $1s$ orbitals $\{\Phi_i({\bf r} )\}$ of an adjustable size $a \equiv \alpha^{-1}$, composing the orthogonalized  atomic (Wannier) functions $\{w_i({\bf r} )\}_{i=1,\ldots ,N}$. The cluster of $N$ atoms with $N$ electrons contains ${2N \choose N }$ states and the second-quantized Hamiltonian is of the form (\ref{ParHam}), with three- and four-site terms added. The three- and four-site interaction terms are difficult to calculate in the Slater basis (see below). Therefore, we have made an ansatz \cite{ThesisRZ} namely, we impose the condition on the trial Wannier function that the three- and four-site matrix elements $V_{ijkl}$ vanish. This allows for an explicit expression of the three- and four- site matrix elements $V'_{ijkl}$ in the atomic representation via the corresponding one- and two- site elements. In Fig. \ref{Fig4} we present the results for the ground- and excited- states energies for the square configuration, $N = 4$ atoms.
The states are grouped into manifolds, which are characterized by the 0,1, and 2 number of double occupancies, appearing in the system. The horizontal lines mark the ground state, states with one and two double electron occupancies in the atomic limit (i.e., for large interatomic distance). The manifolds thus correspond to the {\em Hubbard subbands} introduced for strongly correlated solids \cite{HubbardProcRoySocA1964}. As far as we are aware of, our results are the first manifestation of the energy manifold evolution into well separated subbands with the increasing interatomic distance. The first two subbands correspond to HOMO and LUMO levels
determined in quantum-chemical calculations. In. Fig. \ref{Fig_Phd_Zach} 
we draw the renormalized Wannier function profiles for the $N=6$ atoms.
Note the small negative values on the  nearest-neighbor sites to assure the orthogonality of the functions centered on different 
sites. In the same manner, the electron density profiles can be obtained as a function of intraatomic distances. 

On these examples one can see that both the \emph{ab initio} electronic-structure calculations can be carried out with a simultaneous precise evaluation of microscopic parameters characterizing the particle dynamics and interactions between them. No double counting of the interaction appears at all in either aspect of the calculations. The accuracy of calculating the atomic or molecular structure in the ground state can be reached with accuracy of the order of $1\%$ relatively easy. In the next two Sections extend the method to characterize the Mott physics in nanoscopic one- and two- dimensional systems. 

\section{Mott-Hubbard physics for nanochains: Exact analysis with EDABI} \index{nanochains} \index{Wannier function}

Here we analyze the electronic system properties for for a system composed of $N=6  \div 14$ hydrogen atoms in a linear chain or a ring and draw some universal conclusions about "the nottness". We start with a bit simplified Hamiltonian, but containing  the same principal physics. Namely, 

\begin{equation}
    \hat{\mathcal{H}}= \epsilon_a \sum_{i\sigma}  \hat{n}_{i}
    + t \sum_{i\sigma} \left( \hat{a}_{i\sigma}^{\dag}\, \hat{a}_{i\sigma} +H.c. \right) +
    U\sum_{i}\hat{n}_{i\uparrow}\, \hat{n}_{i\downarrow} + \sum_{i<j}\mathcal{K}_{ij} \hat{n}_{i}\, \hat{n}_{i} +
    \sum_{i<j} V_{ion}(\textbf{r}_{j}-\textbf{r}_{i}).
    \label{Hamiltonian_4}
\end{equation}
The first term represents the atomic energy (we include it explicitly, since $\epsilon_a$ changes with the varying lattice constant). The second describes the kinetic energy of the system with nearest--neighbor
hopping $t$. Next two terms express the intra-- and interatomic Coulomb interaction. The last term is the Coulomb repulsion between the ions located at positions
$\{{\textbf{r}}_i\}$, included for the same reasons as the atomic energy $\epsilon_a$. $V_{ion}$ is the proton-proton classical repulsion term. 

Here we recall only the definitions of single- and
two-particle parameters $t_{ij}$ and $V_{ijkl}$, which are
\begin{equation}
\label{epstdef}
  \epsilon_a=t_{ii}=\langle w_i T w_i\rangle,  \ \ \ \ \ \ \ \
  t=t_{i,i+1}=\langle {w_i}T\rangle{w_{i+1}},
\end{equation}
and
\begin{equation}
\label{ukdef}
  U=V_{iiii}=\left\langle{w_iw_i}|V|{w_iw_i}\right\rangle,  \ \ \ \ \ \ \ \
  K_{ij}=V_{ijij}=\left\langle{w_iw_j}|V|{w_iw_j}\right\rangle.
\end{equation}
The operator $T$ represents the full single--particle lattice potential, i.e.,
\begin{equation}
\label{oper:hop}
  T({\textbf{r}})=-\frac{\hbar^2}{2m}\nabla^2
  -\sum_j\frac{e^2}{\left|{\textbf{r}}-{\textbf{r}}_j\right|}
  {\stackrel{\:\mathrm{a.u.}}{=}}
  -\nabla^2-\sum_j\frac{2}{\left|{\textbf{r}}-{\textbf{r}}_j\right|},
\end{equation}
where $\mbox{a.u.}$ means the expression in  atomic units.
$V=e^2/|{\textbf{r}}_1-{\textbf{r}}_2|$ is the usual Coulomb potential
(we do not include any screening by e.g., core electrons as we want to discuss the model situation, but in a rigorous manner). Analogously, the Coulomb repulsion between ions is $V_{\mathrm{ion}}=V(\textbf{r}_1-\textbf{r}_2)$.

The interatomic Coulomb term in the Hamiltonian can be represented as  
\begin{equation}
\begin{split}
  \sum_{i<j}K_{ij} n_i n_j=\sum_{i<j}K_{ij}(n_i-1)(n_j-1)
  -\sum_{i<j}K_{ij}+2N_e\frac{1}{N}\sum_{i<j}K_{ij}
\\
\vspace{-\abovedisplayskip}
  =H_K+N_e\frac{1}{N}\sum_{i<j}K_{ij}+(N_e-N)\frac{1}{N}\sum_{i<j}K_{ij}, 
  \label{hkdef}
  \end{split}
\end{equation}
where we use the relation $N_e=\sum_in_i$ and introduce the symbol $H_K$ for the longer--range Coulomb interaction.
Substituting (\ref{hkdef}) into (\ref{Hamiltonian_4}) and representing the ionic repulsion in the form
$$
  \sum_{i<j}\frac{2}{R_{ij}}=N_e\frac{1}{N}\sum_{i<j}\frac{2}{R_{ij}}
  -(N_e-N)\frac{1}{N}\sum_{i<j}\frac{2}{R_{ij}}
$$
(in Rydbergs), where $R_{ij}=|{\textbf{r}_j-\textbf{r}_i}|$, we obtain that
\begin{equation}
\label{fulhameff}
  H=N_e\epsilon_a^\mathrm{eff}+H_t+H_U+H_K+(N_e-N)\frac{1}{N}\sum_{i<j}
  \left(K_{ij}-\frac{2}{R_{ij}}\right),
\end{equation}
where the kinetic energy and intraatomic Coulomb interaction terms are $H_t$ and $H_U$, and the \emph{effective atomic energy} is defined (in Ry) as
\begin{equation}
\label{eaeff}
  \epsilon_a^{\mathrm{eff}} \equiv \epsilon_a+\frac{1}{N}\sum_{i<j}
  \left(K_{ij}+\frac{2}{R_{ij}}\right).
\end{equation}
The effective atomic energy contains the electron attraction to the ions, as well as the mean--field part of their repulsion ($K_{ij}$), and the ion--ion interaction. Such a definition preserves correctly the atomic limit, when the distant atoms
should be regarded as neutral objects. In practice, the above form is calculated numerically with the help of Richardson extrapolation for $N\rightarrow\infty$.
One can find it converges exponentially with $N$, whereas \emph{bare} $\epsilon_a$ is divergent harmonically, due to $\sim 1/r$ Coulomb wells in the single--particle potential (\ref{oper:hop}).

The last term in the Hamiltonian (\ref{fulhameff}) vanishes for the half--filled band case $N_e=N$. It also does not affect the system charge gap (as it depends linearly on $N_e$), and the correlation functions away from half filling.
Therefore, we can write down the system Hamiltonian in the more compact form

\begin{equation}
\label{hameff}
  H=\epsilon_a^\mathrm{eff}\sum_i n_i
  + t\sum_{i\sigma}\left(a_{i\sigma}^{\dagger}a_{i+1\sigma}+\mbox{HC}\right)
  + U\sum_in_{i\uparrow}n_{i\downarrow} + \sum_{i<j}K_{ij}{\delta n_i}{\delta n_j},
\end{equation}
where $\delta n_{i}\equiv n_i-1$. Thus, all the \emph{mean--field} Coulomb terms are collected into $\epsilon_a^\mathrm{eff}$. 

In the framework of \emph{tight--binding approximation} (TBA) one can postulate Wannier functions in a simple form, which is validated by an exponential drop of Wannier functions.
The orthogonality relation $\langle{w_i|w_{i\pm 1}}\rangle=0$ and the normalization
condition $\langle{w_i|w_i}\rangle=1$ leads to coefficients of the expansion
\begin{equation}
\label{wannga}
  \gamma=\frac{S_1}{(1+S_2)+\sqrt{(1+S_2)^2-S_1(3S_1+S_3)}},
\end{equation}
and
\begin{equation}
\label{wannbe}
  \beta=\left[1-4\gamma S_1+2\gamma^2(1+S_2)\right]^{-1/2},
\end{equation}
where we define the overlap integral of atomic functions $S_m=\langle{\Psi_i|\Psi_{i+m}}\rangle$ (the normalization $S_0=\langle{\Psi_i|\Psi_i}\rangle=1$ is assumed). The above expressions are well--defined if the quantity under the square
root of Eq.\ (\ref{wannga})
\begin{equation}
  \Delta \equiv (1+S_2)^2-S_1(3S_1+S_3)>0,    
\end{equation}
(cf. Table \ref{wann:tab}). The actual limits of TBA comes with nonzero overlap integral of Wannier functions, 
when including the second--neighbor contribution (see Table \ref{wann:tab}), i.e.,
$$
  \langle{w_i|w_{i+2}}\rangle=\beta^2\gamma^2.
$$
The above nonorthogonality may strongly affect the second neighbor hopping,
as a zero--overlap is crucial for the convergence of hopping integral on
an lattice providing the single--particle potential of the form
(\ref{oper:hop}). However, as the only term involving second--neighbors in our Hamiltonian (\ref{hameff}) is the interatomic Coulomb repulsion $K_2$, the presented TBA approach seems sufficient for the purpose (for details see \cite{RycerzPhDThesis}). \index{Gaussian STO-3G basis}

\begin{table}[!t]
\caption{Wannier--basis parameters for 1D chain calculated in the Gaussian STO--3G basis, with adjustable size, as a function of lattice parameter $a$ ($a_0$ is the Bohr radius).
The values of the optimal inverse orbital size $\alpha_{\min}$ are also
provided.}
\label{wann:tab}
\begin{center}
\begin{tabular}{rcllll}
\hline\hline
$a/a_0$ & ${\alpha}_{\min}a_0$ & $\beta$ & $\gamma$ &
$\left<w_i|w_{i+2}\right>$ & $\Delta$ \\ \hline
 1.5 & 1.363 & 1.41984 & 0.32800 & 0.21689 &      0.34735 \\
 2.0 & 1.220 & 1.23731 & 0.26301 & 0.10590 &      0.50525 \\
 2.5 & 1.122 & 1.14133 & 0.20965 & 0.05725 &     0.63980 \\
 3.0 & 1.062 & 1.08190 & 0.16246 & 0.03089 &      0.75691 \\
 3.5 & 1.031 & 1.04394 & 0.12013 & 0.01573 &     0.85349 \\
 4.0 & 1.013 & 1.02216 & 0.08568 & 0.00768 &   0.92009 \\
 4.5 & 1.007 & 1.01010 & 0.05795 & 0.00343 &    0.96170 \\
 5.0 & 1.004 & 1.00429 & 0.03779 & 0.00144 &   0.98327 \\
 6.0 & 1.001 & 1.00063 & 0.01451 & 0.00021 &   0.99749 \\
 7.0 & 1.000 & 1.00007 & 0.00483 & 2.3$\cdot 10^{-5}$ & 0.99972 \\
 8.0 & 1.000 & 1.00001 & 0.00139 & 1.9$\cdot 10^{-6}$ & 0.99998 \\
10.0 & 1.000 & 1       & 7.3$\cdot 10^{-5}$ & 5.3$\cdot 10^{-9}$ & 1 \\
\hline\hline
\end{tabular}
\end{center}
\end{table}

We already mentioned, that the atomic energy $\epsilon_a$ is divergent with the
lattice size $N$ and define the convergent effective quantity $\epsilon_a^\mathrm{eff}$
(\ref{eaeff}). 

\begin{table}[!b]
\caption{Microscopic parameters (in Ry) of 1D chain, calculated
in the Gaussian STO--3G basis. Corresponding values of the optimal inverse
orbital size $\alpha_{\rm min}$ are provided in Table \ref{wann:tab}.
The Richardson extrapolation with $N\rightarrow\infty$ were used.}
\label{tuvsk:tab}
\vspace{1em}
\begin{tabular}{rcllrrllll}
\hline\hline
$R/a_0$ & $\epsilon_a^\mathrm{eff}$ & $t$ & $U$
 & $V^{\dagger}$ & $J^{\dagger}$ & $K_1$ & $K_2$ & $K_3$ \\
\hline
1.5 &  0.0997 & -0.8309 & 2.054
 & -43.93 & 30.92 &   1.165 & 0.667 & 0.447  \\
2.0 &  -0.5495 & -0.4423 & 1.733
 & -23.81 & 21.06 &   0.911 & 0.501 & 0.334  \\
2.5 &  -0.7973 & -0.2644 & 1.531
 & -14.95 & 15.13 &   0.750 & 0.401 & 0.267  \\
3.0 &  -0.9015 & -0.1708 & 1.407
 & -10.99 & 10.91 &   0.639 & 0.334 & 0.222  \\
3.5 & -0.9483 & -0.1156
 & 1.335 & -9.41 & 75.6 & 0.557 & 0.286 & 0.191 \\
4.0 &  -0.9705 & -0.0796 & 1.291
 & -8.74 & 4.93 &   0.493 & 0.250 & 0.167  \\
4.5 & -0.9815 & -0.0549
 & 1.270 & -8.10 & 2.92 & 0.442 & 0.222 & 0.148  \\
5.0 &  -0.9869 & -0.0374 & 1.258
 & -7.07 & 1.57 &   0.399 & 0.200 & 0.133  \\
6.0 & -0.9908 & -0.01676
 & 1.249 & -4.29 & 0.34 & 0.333 & 0.167 & 0.111 \\
7.0 & -0.9915 & -0.00710
 & 1.247 & -1.96 & 0.05 & 0.286 & 0.146 & 0.095 \\
8.0 &  -0.9917 & -0.0027 & 1.247
 & -0.70 & 5$\cdot 10^{-3}$   & 0.250 & 0.125 & 0.083  \\
10.0 &  -0.9917 & -2.5$\cdot 10^{-3}$ & 1.247
 & -0.05 & 2$\cdot 10^{-5}$   & 0.200 & 0.100 & 0.067  \\
\hline\hline
\end{tabular}

$^{\dagger}$ The values of $V$ and $J$ are specified in mRy.
\end{table}

The values of the model parameters, corresponding to the lattice spacing
$a/a_0=1.5\div 10$, are presented in Table \ref{tuvsk:tab}.
The data correspond to the optimal values of the inverse orbital size
$\alpha_{\rm min}$, as displayed in Table \ref{wann:tab}.
We also provide there the values of the \emph{correlated hopping} $V$ and the
\emph{Heisenberg--exchange integral} $J$ to show that one could indeed disregard the corresponding terms in the Hamiltonian (\ref{hameff}).

One can note the values of $t$ calculated in the Gaussian STO--3G basis (listed in Table \ref{tuvsk:tab})
differs from those obtained in Slater basis 
by less then $0.5\%$ when using the same values of the inverse orbital size
$\alpha$.
However, the differences grow significantly, if $\alpha$ is optimized
independently for the Slater basis and the three-- and four--site terms
are not included in the atomic basis.

\subsection{Results}
\label{optigsen}
We now consider a nanoscopic linear chain of $N=6\div 14$ atoms,
each containing a single valence electron (hydrogenic--like atoms),
including \emph{all} long-range Coulomb interactions
(3-- and 4--site terms are also included in the adjustable Gaussian  STO--3G basis).

\subsubsection{Crossover from metallic to Mott-Hubbard regime}
\label{edabires}

The Hamiltonian (59) is diagonalized in the Fock space with the help of Lanczos method. As the microscopic parameters $\epsilon_a^\mathrm{eff}$, $t$, $U$, and $K_{ij}$ are calculated numerically
in the Gaussian STO--3G basis, the inverse orbital size $\alpha$ of the 1$s$--like state is subsequently optimized to obtain the ground state
energy $E_G$ as a function of the interatomic distance $a$. \index{Mott-Hubard}

\begin{figure} 
    \centering
   \includegraphics[width=0.6\textwidth]{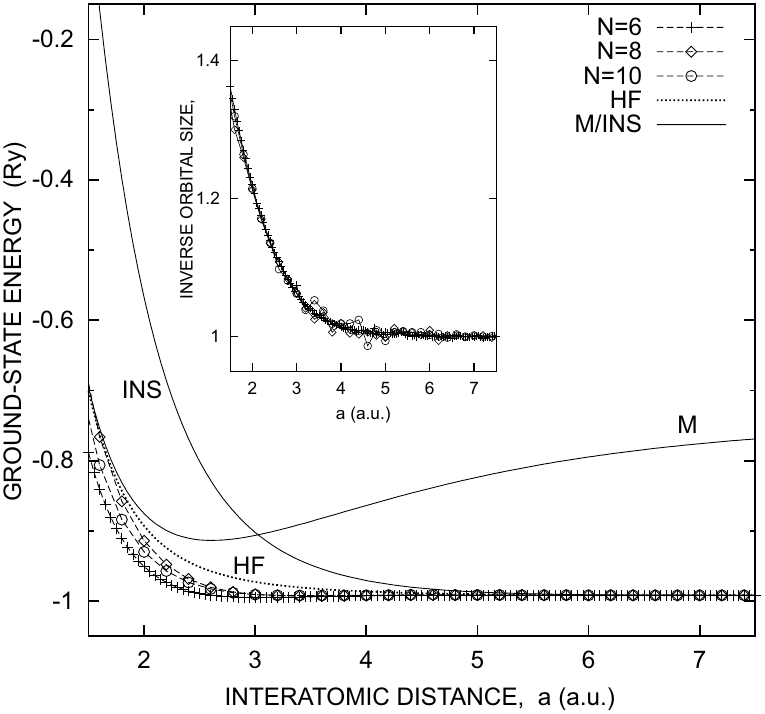}.
    \caption{The ground state energy per atom for the linear chain of $N=6\div 10$ atoms with periodic boundary conditions. The Gaussian-type orbitals (STO--3G basis) with their adjustable inverse size $\alpha^{-1}$ have been used. The energies of the {\em ideal} metallic (M),
\emph{ideal} insulating (INS), and Hartree--Fock (HF) solutions for an
\emph{infinite} system are shown for comparison. The {\em inset} provides the
optimal inverse orbital size $\alpha_{\min}$.}
\label{egedabi}
\end{figure}

Their effects on convergence of the results for the ground-state energy
$E_G$ and the optimal inverse orbital size $\alpha_{\rm min}$ are shown in
Figure \ref{egedabi} for $N=6\div 10$ atoms.
These results were used to extra\-po\-la\-te the value of the variational
parameter $\alpha_{\rm min}$ to larger $N$ to speed up the computations.
Figure \ref{egedabi} illustrates also the \emph{Mott-Hubbard localization criterion.} 
Namely, for the interatomic distance $a\approx 3a_0$ the energy of the \emph{ideal metallic} state (M), determined as

\begin{equation}
\label{egmet}
  E_G^{\rm M}=\epsilon_a^\eff-\frac{4}{\pi}\abs{t} + \frac{1}{N}
  \sum_{i<j}K_{ij}\aver{\delta n_i\delta n_j},
\end{equation}
where the charge--density correlation function $\langle{\de n_i\de n_j}\rangle$ is taken for 1D electron-gas on the lattice
\begin{equation}
   \aver{\delta n_i\delta n_j}=
  -2\frac{\sin^2(\pi\abs{i\!-\!\!j}/2)}{(\pi\abs{i\!-\!\!j})^2}
\end{equation}
(for the \emph{half--filled} band case), crosses over to that representing the \emph{Mott insulating} state (INS), with
\begin{equation}
\label{egins}
  E_G^{\rm INS}=\ep_a^\eff.
\end{equation}
One usually adds the second--order perturbation correction to the energy of insulating state (\ref{egins}) in the well--known form \cite{SpalekPhysStatSol1981}
\begin{equation}
  \frac{4t^2}{U-K_1}\left(\aver{\mathbf{S}_i\cdot\mathbf{S}_{i+1}}   -\frac{1}{4}\right),
\end{equation}
the Bethe--Ansatz result is
$\aver{\mathbf{S}_i\cdot\mathbf{S}_{i+1}}-1/4=-\ln 2$ for the quantum Heisenberg
antiferromagnet.
Here we only compare the two simplest  approaches, leading to the
energies (\ref{egmet}) and (\ref{egins}).

The critical value of $a$ is very close to obtained for the 1$s$ Slater--type orbitals.
The validity of the above Mott-Hubbard criterion for this one--dimensional system is quantitative, as the energy of the antiferromagnetic (so \emph{Slater-type}) Hartree--Fock solution (HF) is lower than those of the paramagnetic M and INS states.
Therefore, a detailed verification of this criterion would be estimating the charge--energy gap and transport properties of this correlated system directly.

\subsubsection{Evolution of Fermi-Dirac statistical distribution into continuous spread of localized states}

We consider now on the principal and exact results for model linear chains of hydrogen atoms. Those results can be viewed as  concerning quantum monoatomic nano-wires composed of elements with one valence electrons, and the inner-shell electrons treated as part of ionic case. Before presenting the physical properties we characterize briefly the methodology of our approach. First, the single-particle basis is selected in such a way that each of the starting atomic wave function (Slater orbital) is composed of three Gaussians STO-3G of adjustable size. Out of them one constructs the Wannier basis with the help of which we determine the trial values of the microscopic parameters of Hamiltonian (\ref{hameff}).
Second, in accordance with the scheme presented in Fig. 7, we diagonalize the Hamiltonian in the Fock space for $N\leqslant 14$ atoms using the Lanczos algorithm. At the end, we optimize the orbital size $\alpha^{-1}$ and thus the results can be presented as a function of interatomic distance, which mimics the gradual transformation of collective (itinerant)  states at small distances into a set of atomic states, emplifying the Mott-Hubbard insulator. 

\begin{figure} 
    \centering
   \includegraphics[width=0.9\textwidth]{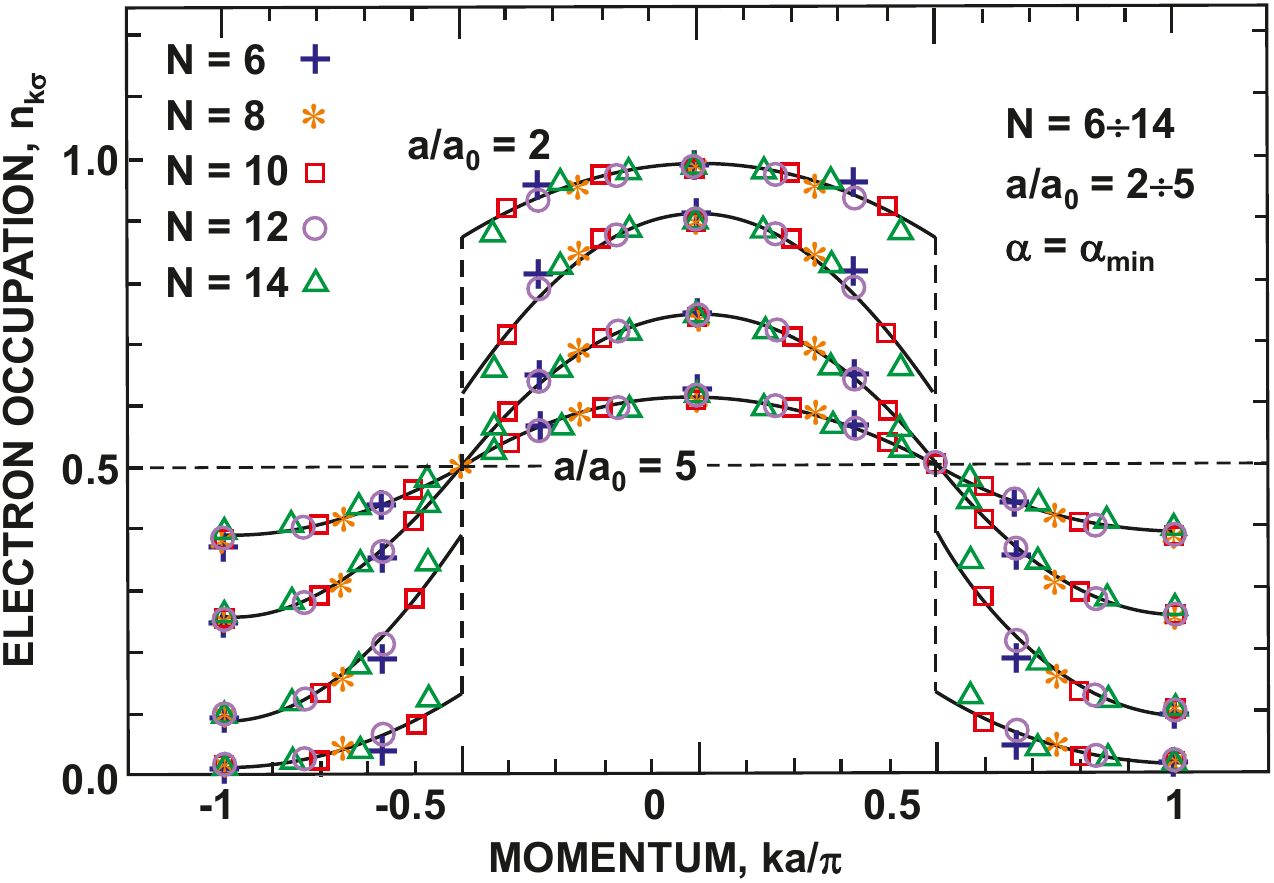}.
    \caption{Momentum distribution for nanochain of interacting $N$ hydrogen atoms.  The distribution is smeared out with the increasing lattice constant $a$ and reaches its average $\simeq 0.5$ for the critical distance $a\simeq 5a_{0}$. The threshold value of $a$ signals a crossover transformation to the localized state.}
\label{nksqf}
\end{figure}

Few words about the modified boundary conditions should be added. We take the periodic conditions for the systems with $N=4n+2$ atoms and antiperiodic for $N=4n+4$. In the case of odd $N$ the phase is defined with value between the above two cases, where the wavefunction phase changes $\phi$ by $2\pi$ and $\pi$, respectively). In Fig. \ref{nksqf} we present the statistical distribution function $n_{k\sigma}$ for $N=6\div 14$ atoms in the chain. This is one of our principal results. The solid state lines represent a singular polynomial fit \cite{Solyom}
\begin{equation}
    n_{k\sigma}=n_{F}+A\left|k_{F}-k\right|^{\Theta} \mbox{sgn}\left(k-k_{F}\right), 
\end{equation}
with a non-universal (interaction dependent) exponent $\Theta$ ranging from $0.4$ (for $a=2a_0$) t $\Theta\simeq 1.5$ for $a\gtrsim 4a_0$. Also, a finite jump of $\Delta n_F$ is observed at the Fermi momentum providing a quasi-particle type normalization factor $Z_{k_{F}}$.

\begin{figure} 
    \centering
   \includegraphics[width=0.9\textwidth]{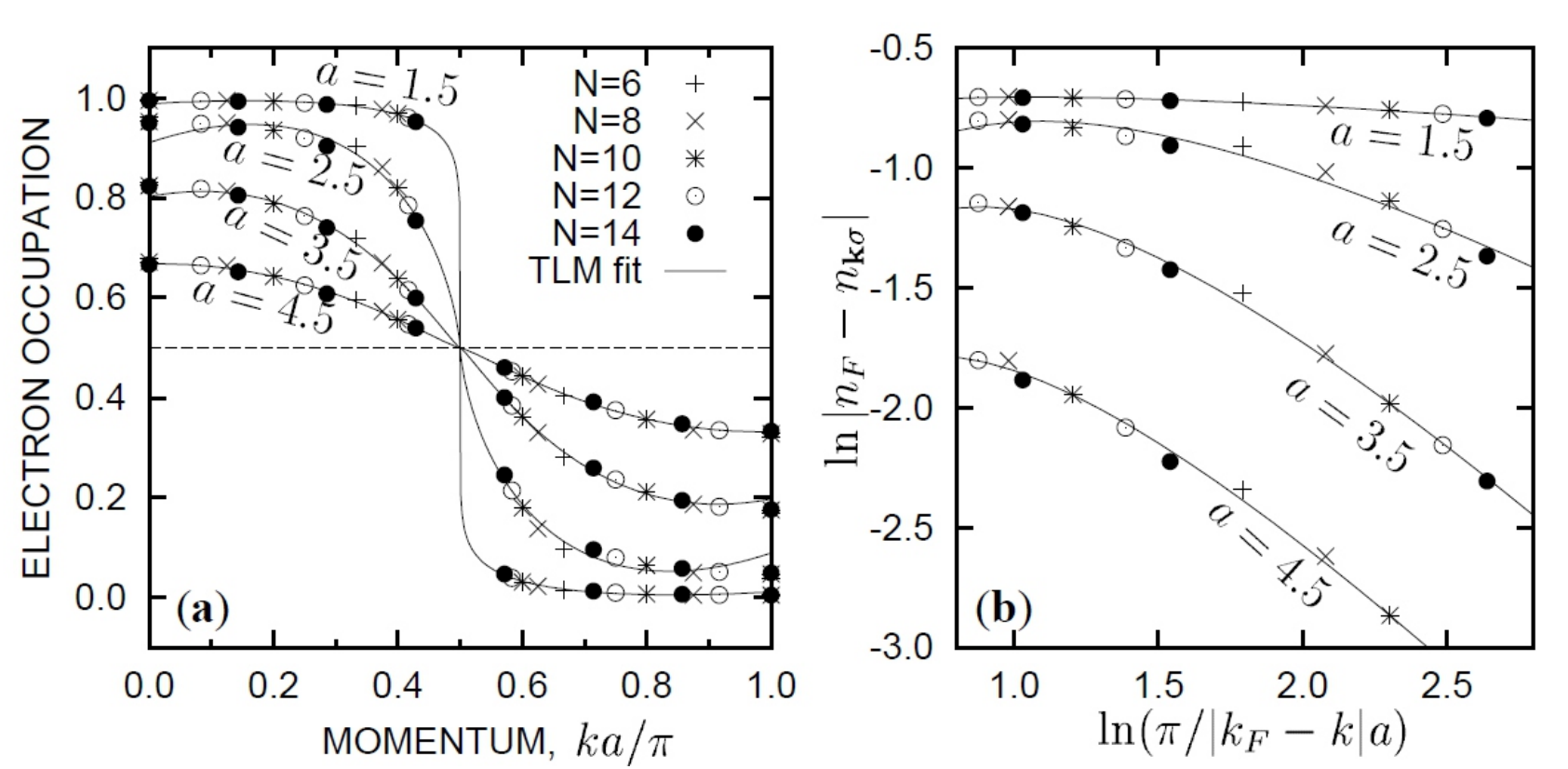}.
    \caption{Tomonaga--Luttinger--liquid scaling for a \emph{half--filled} nano-chain of
$N=6\div 14$ atoms with \emph{long--range} Coulomb interactions:
$(a)$ momentum distribution in the linear and $(b)$ the same in the log--log
scale: continuous lines represent the fitted singular expansion in powers of
$\ln(\pi/|k_F-k|a)$}
\label{nksll}
\end{figure}

The situation of this nano-Fermi liquid can be characterized equally well by Tomonaga-Luttinger-model scaling (TLM) scaling \cite{RycerzPhDThesis}  of this nanoliquid  depicted is Fig. \ref{nksll}. In conclusion, although the two types of fitting procedures work almost equally well, the intermediate character of the nanoliquid between the Fermi and T-L limits has the value of its own.

\begin{figure}[ht!]  
\centering
   \includegraphics[width=0.9\textwidth]{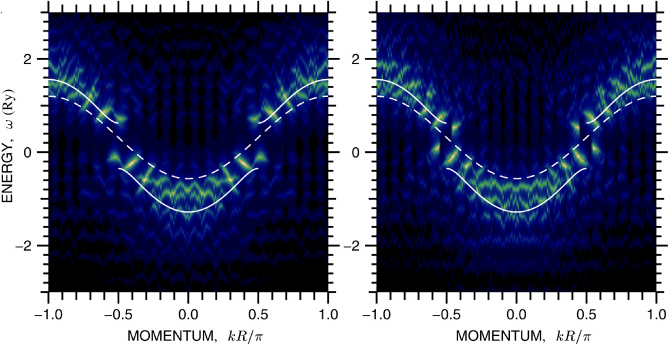}
\caption{Exemplary electronic structure of electrons in a nanochain of $N = 10$ (left panel) and $N = 11$ (right panel) atoms as a representation of spectral-density-peak positions. The solid lines represent the Hartree-Fock  results and the dashed lines represent the result for noninteracting electrons. 
For explanation of the gap see main text.}
\label{Fig_15}
\end{figure}

A direct demonstration of the emerging electronic structure is exemplified in Fig. 15, which was obtained by calculating from the definition of the spectral density. The result is shown in Fig. \ref{Fig_15}  
for $N=10$ (left panel) and $N=11$ (right panel). The main novel feature is the splitting accusing at the nominal Fermi-momentum points. The solid line is the calculated electronic structure in the Hartree-Fock approximation and the dashed line is the band structure for noninteracting electrons with $t$ values obtained from EDABI. The striking feature of this electronic structure is the splitting occuring at $k_F=\pm \pi/a$ signaling onset of the antiferromagnetic superstructureappearing even for this very small system in the ground state. An explanation of this surprising feature emerges directly from the calculations of the spin-spin correlation function $\langle \pmb{S}_i \cdot \pmb{S}_j \rangle$, as illustrated in Fig. 16. We see that the correlations persist throughout the whole system length. In such circumstance the system behaves as if it possessed a long-range order, a truly collective behavior of a system with $N_e\sim 10$ electrons, but with long-range interaction -  Coulomb interaction included.  

\begin{figure}[ht!]
\centering
   \includegraphics[width=\textwidth]{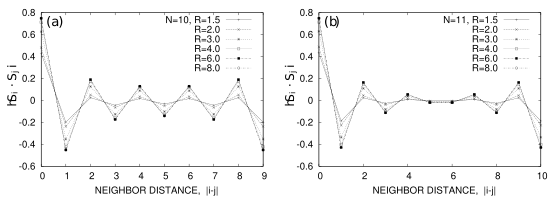}
\caption{Parity effect on spin ordering: spin--spin correlations
  for nanochains of $N=10$ $(a)$ and $N=11$ $(b)$ atoms. The values
  of the interatomic distance $R$ are specified in the atomic units
  ($a_0=0.529\mbox{ \AA}$).}
\label{ssfig}
\end{figure}

\section{Recent developments}

Here we would like to mention application of EDABI method to the problem of metalization of molecular hydrogen. First, the insulator-metal transition (hydrogen metalization of the insulating molecular hydrogen into metal) is a discontinuous transition from a diamagnet into paramagnet, so this transition should be describable in our local language. Second, we have extend the cluster EDABI analysis to a bulk system, both in one- and two-dimensions. Here we present only the results for the latter situation. \index{EDABI method}

In Fig. 17 we the present schematically \ce{H_2} molecules stacked vertically and forming square lattice, which is divided into super-cells sp       ecified explicitly there. The extent Hubbard Hamiltonian of a single supercell is diagonalized exactly, and repeated periodically, with the intermolecular Coulomb interaction between the cells included (for details see \cite{Biborski}). The system enthalpy is calculated as a function of pressure and the relevant renormalized Wannier functions $\{w_{i}^{u}(\textbf{r})\}$, with $\mu=1,2$ characterizing  the adjusted wave functions of the individual $1s$ is atomic states in the molecule, are determinated in the whole procedure. As a result, the phase diagram on the plane pressure entalphy is determined and comprises molecular - molecular and molecular - atomic solid discontinuous phase transitions. This phase diagram is drawn in Fig. 18. Those results may serve as a starting point to a more comprehensive analysis of complex 
face diagram of solid hydrogen. The molecular to atomic solid transition can be characterized as an example of the Mott-Hubbard transition, albeit from molecular solid to a metal \cite{Biborski}. In brief, this example also shows that the EDABI method may be applied to real systems and to the localization-delocalization transition of a nonstandard nature. 
As the most recent example of the EDABI application we display the total energy of \ce{LiH} and \ce{LiH\cdot H_2} clusters (M. Hendzel, private communication) within an extended basis involving starting $1s$, $2s$, and $2p$ orbitals of variable size. The values are close to the experimental values for those systems are: -16.1611 and -17.8942 Ry, respectively. We should see progress along these lines in the near future. 

\begin{figure}[h] 
\centering
   \includegraphics[width=0.9\textwidth]{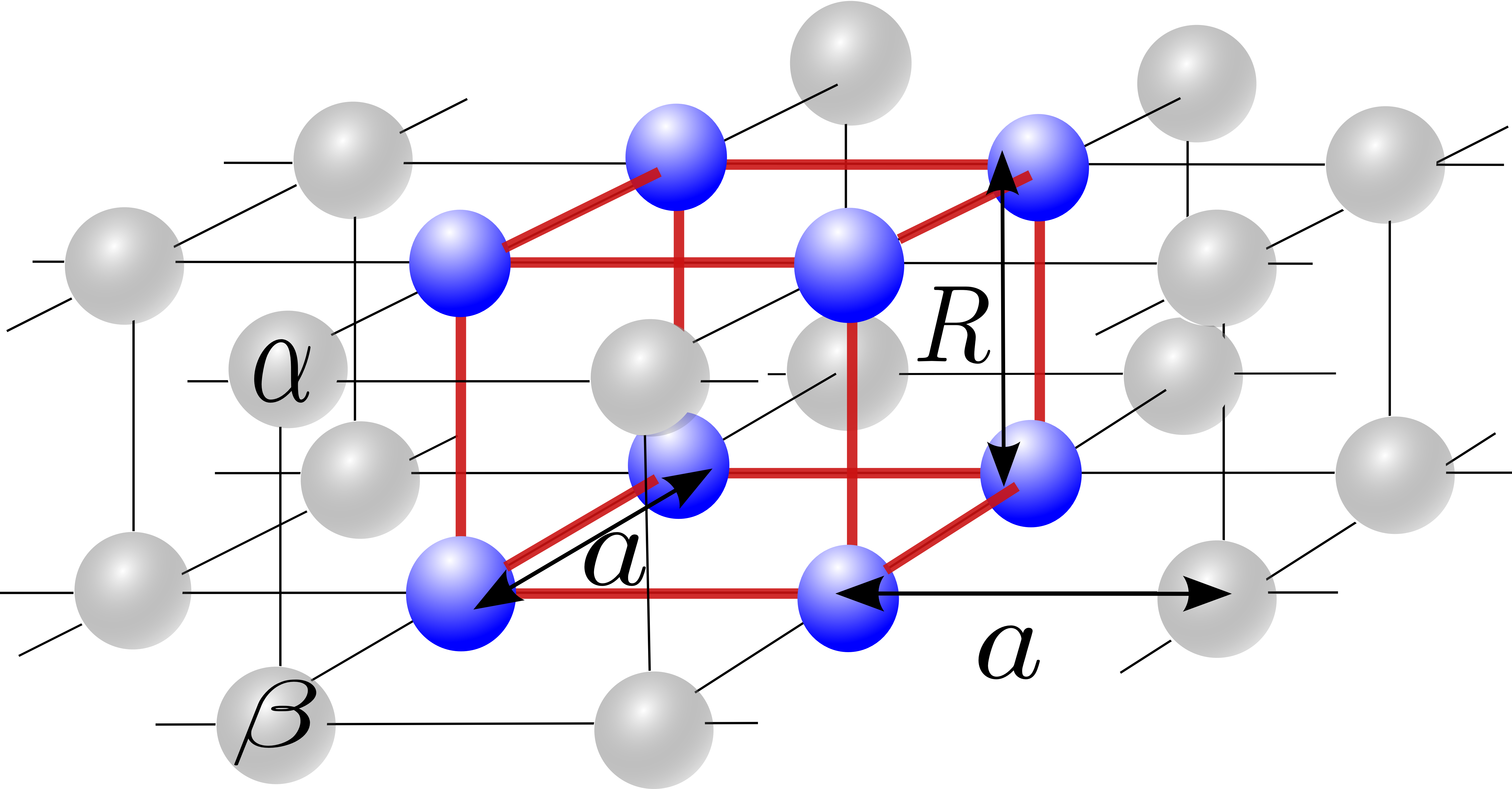}
 \caption{Schematic representation of stacked vertically $H_2$ molecular $2D$ layer forming square lattice. The bond length and the intermolecular distance are marked by $R$ and $a$, respectively. There are eight
 atoms in the supercell (dark blue spheres). The supercell is repeated periodically to conform periodic boundary conditions (PBC). Shaded spheres indicate atoms which are continuations resulting from the PBC implementation. The indicies $\alpha$, $\beta$ distinguish the component atoms of each molecule.}
 \label{fig:plane_model}
\end{figure}

\begin{figure}[h] 
\centering
   \includegraphics[width=0.8\textwidth]{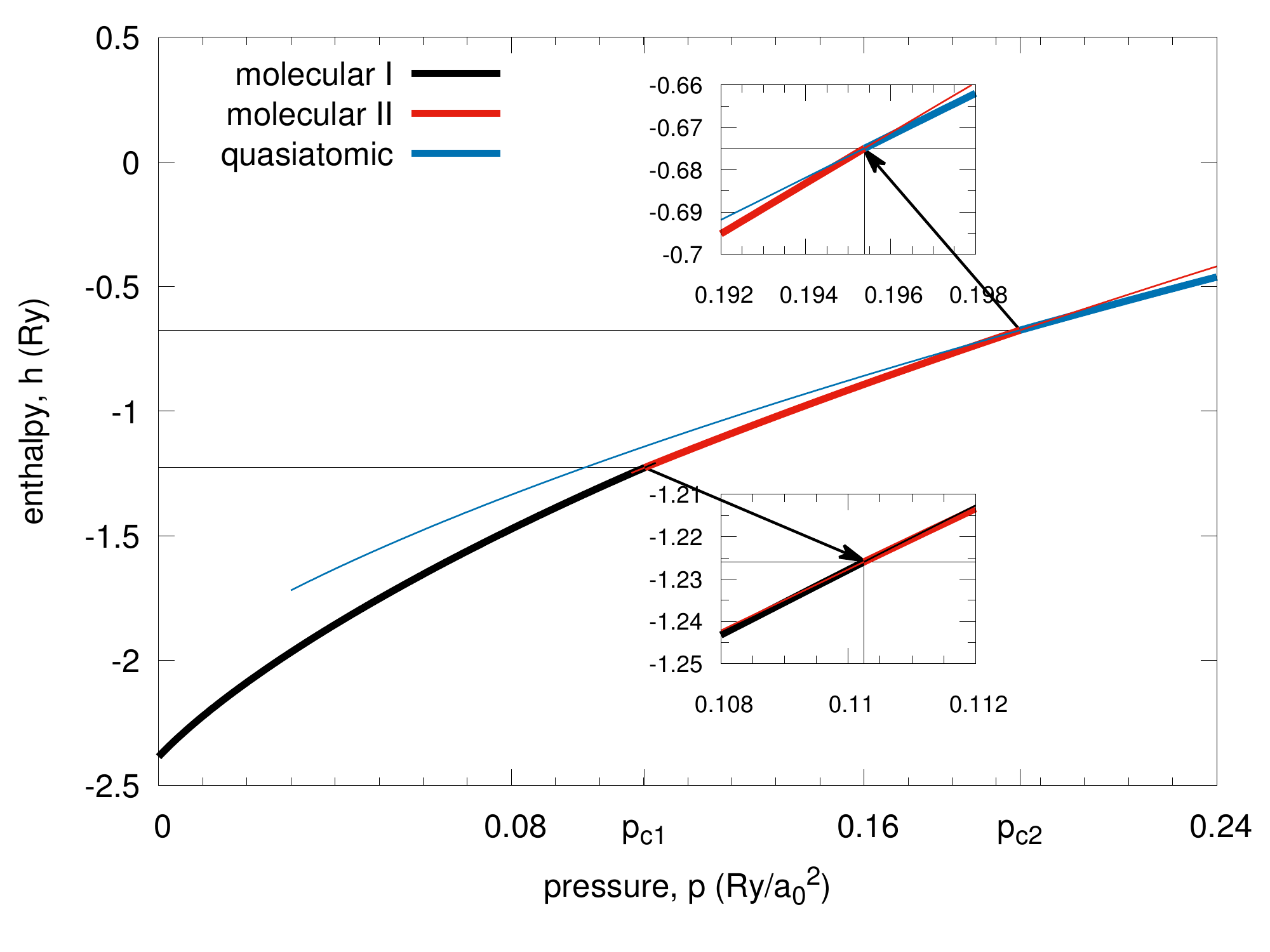}
 \caption{The enthalpy (per molecule) versus pressure $p$. At lower pressure, two molecular phases are stable; the transition to the \emph{quasiatomic} phase
 occurs at $p_{c2} \sim 0.1954 Ry/a_{0}^2$, as marked. $E_B(p=0)=-2.3858 Ry$, $R_{eff}(p=0) = 1.4031 a_0$, $a(p=0) = 4.3371 a_0$. Thin lines extrapolate the enthalpies of the particular phases beyond the regime of their stability. Insets show some detail of the transitions.}
 \label{fig:2d_enthalpy}
\end{figure}

\clearpage
\newpage
\begin{figure}[ht!] 
\centering
   \includegraphics[width=0.9\textwidth]{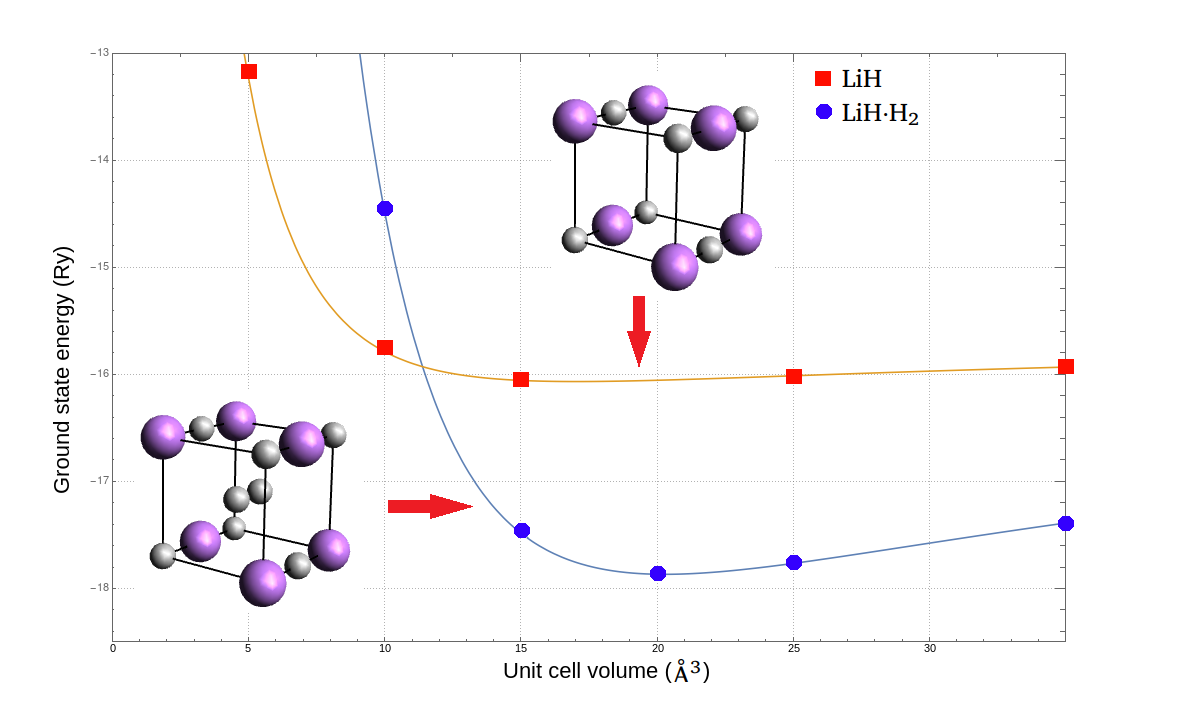}
 \caption{Total energy for \ce{LiH} (squares) and \ce{LiH\cdot H_2} clusters as a function of unit cell volume as obtained from EDABI
 [M. Hendzel, unpublished]}
 \label{lih}
\end{figure}

\section{Conclusions}

The {\bf E}xact {\bf D}iagonalization \emph{\textbf{A}b {\bf I}nito} (EDABI) method has been developing slowly and so far is useful mainly in describing in a precise manner model systems. Its principle aim has been to determine properties of correlated systems by incorporating also the calculations of model parameters, such as the Hubbard U, the hopping integrals $t_{ij}$, etc., into the general scheme of electronic structure calculations. It is particularly well suited for a description of nanosystems, in which a crossover from atomic to itinerant character of electronic states occurs a sign of the Mott-Hubbard behavior. The studied evolution with the lattice parameter emulates the pressure dependence of the basic quantum properties and correlation functions. On the examples discussed in Sections 4 and 5  collective 
(bulk) properties are exhibited in a direct manner. The future studies should show to what extent can the results be analyzed experimentally. 
Finally, we summarize the fundamental features of the EDABI method:
\begin{enumerate}
  \item [\emph{1}$^{\circ}$] The 1$^{\rm{st}}$ and 2$^{\rm{nd}}$ quantization aspects of the collective (nano)systems are tackled in a consistent manner, i.e., without encountering the problem of double counting interactions, as is the case in the present versions of DFT + U and DFT + DMFT treatments.
 \item [\emph{2}$^{\circ}$] In the approach we first diagonalize second-quantized Hamiltonian for selected trial single- particle wavefunction basis and optimize it subsequently in the correlated state. In other words, the usual quantum-mechanical procedure of determining their e.g., the system energy is carried out in a reverse order (the correlations are as crucial as the single-particle wave function evaluation). 
\item [\emph{3}$^{\circ}$] 
 The method allows to analyze within a single scheme atomic, molecular, and extended systems via studies of nanoscopic systems of the increasing size.
 
 Progress in calculating precisely properties of collective systems composed of more complex atoms will be effective within is method only with the implementation of computing capabilities, perhaps coming with the advent advanced quantum computing.
\end{enumerate}

I am grateful to my former students: Adam Rycerz, Edward M G\"{o}rlich, Roman Zahorbe\'{n}ski, and Andrzej Kadzielawa, as well as to Dr. Andrzej 
Biborski, for permission to use some of the numerical results from their Ph.D. Theses and material from joint publications. I would like to thank Dr. Danuta Goc-Jag{\l}o and Dr. Maciej Fidrysiak for their technical help with formatting the text and figures. Mr Maciej Hendzel provided Fig. \ref{lih}, coming from his unpublished M. Sc. Thesis.
The work here was financed through the Grant OPUS No.UMO-2018/29/B/ST3/02646 from Narodowe Centrum Nauki (NCN).
\bibliography{bibliography}


\end{document}